%% file: MainDoc.tex
\documentclass[12pt, oneside]{article} 
\usepackage[left=2.5cm,top=2.5cm,right=2.5cm,bottom=2.5cm]{geometry}
\usepackage[utf8]{inputenc} 
\usepackage{latexsym} 
\usepackage{amsmath} 
\usepackage{amssymb} 
\usepackage{float}
\usepackage{amsthm} 
\usepackage{graphicx} 
\newcommand\fnote[1]{\captionsetup{font=small}\caption*{#1}}
\usepackage{epsfig}
\usepackage[font=small,labelfont=bf]{caption}
\usepackage{sectsty}
\usepackage{setspace}
\usepackage{natbib}
\usepackage{hyperref}
\usepackage{caption}
\usepackage{subcaption}
\usepackage{lscape} 
\usepackage{booktabs,caption}
\usepackage[flushleft]{threeparttable}
\usepackage[referable]{threeparttablex}

\hypersetup{
  colorlinks   = true,
  urlcolor     = blue,
  linkcolor    = blue,
  citecolor   =  blue,
}

\title{\textbf{The unintended effects of universalizing social pensions: Evidence from Mexico}\vspace{-9mm}}

\author{
  Oscar Galvez-Soriano\footnote{Corresponding Author. Department of Economics,  University of Chicago, Saieh Hall for Economics, Chicago, IL 60637. e-mail: \href{mailto: ogalvez@uchicago.edu}{ogalvez@uchicago.edu}}\\
  The University of Chicago
   \and
  Raymundo Ramirez Peralta\footnote{Department of Economics, El Colegio de Mexico, Carretera Picacho Ajusco 20 Ampliación, Mexico City, CDMX 14110.}\\
  El Colegio de Mexico
\vspace{-2mm}
}

\date{\today \vspace{-12mm}}

\begin{document}
\maketitle

\begin{abstract}
We examine the effects of the 2019 expansion of Mexico’s Social Pension Program. This reform simultaneously increased benefit generosity and expanded eligibility to individuals previously excluded because they received contributory pensions. Using nationally representative household data and a combination of difference-in-differences and triple-differences strategies, we separately identify the effects of increased transfer generosity and expanded eligibility. We find that the combined reform substantially increased program participation, household income, and reduced both poverty and extreme poverty. However, these gains were driven primarily by the increase in benefit levels. When we isolate the effect of universalization, we find no significant reduction in poverty and a modest increase in extreme poverty by 2024. Consistent with a labor-leisure framework, this increase is concentrated among low-income elderly individuals who reduced labor supply after receiving the pension despite the transfer being insufficient to fully replace forgone earnings. The universalization also expanded participation among low-income elderly who may have previously faced barriers to enrollment while attracting substantial participation among higher-income beneficiaries whose economic behavior remained largely unchanged. In addition, the reform generated heterogeneous effects on household consumption, health expenditures, and educational investments. Overall, the results highlight an important trade-off of universal social pensions: broader access can reduce exclusion, but the welfare consequences of expanding coverage may differ substantially from those of increasing transfer generosity.

\end{abstract}

\begin{description}
\item [{JEL Classification:}] H55, I38, J14, O15. \vspace{-3mm}
\item [{Keywords:}] Non-contributory pensions, social protection, retirement behavior, poverty.
\end{description}

\newpage{}

\doublespacing



\section*{Introduction}
In the face of demographic aging and limited social security coverage, non-contributory pensions have emerged as a critical policy tool for improving the well-being of older adults in low- and middle-income countries. Designed to provide regular cash support to elderly individuals excluded from formal retirement systems, these programs have become a central pillar of social protection strategies, particularly in regions with high rates of old-age poverty and labor market informality. A growing body of research has emphasized their potential to enhance consumption, reduce vulnerability, and influence household dynamics, especially where traditional family support structures are weakening or insufficient to meet the needs of aging populations \citep{barrientos2006, Case2007, leisering2009, dethier2010, aguila2020living}.

Initial studies of social pensions focused on their effects on household consumption, education, and intra-household resource allocation, rather than directly on poverty outcomes \citep{case1998, Duflo2003, aguila2017}. This focus partly reflected the fact that early programs were typically means-tested and thus targeted to the poorest, limiting the scope for analyzing marginal poverty effects. However, even when not means-tested, as in some rural programs in China, India, and Mexico, much of the empirical literature prioritized labor market behavior \citep{ardington2009, juarez2015} and select health indicators, such as depression and subjective well-being \citep{Kaushal2014, Galiani2016, huang2021}.

More recently, a second generation of research has begun to systematically evaluate food security and the poverty-reducing potential of these programs. Newer studies of Bolivia, China, Hong Kong, India, and Mexico find that non-contributory pensions significantly reduce extreme poverty, although their effects on broader poverty measures and labor force participation are more mixed \citep{Unnikrishnan2020, Bottan2021, avila2024, Zhu2023, zhou2025}. These contributions have also highlighted important policy trade-offs, including the extent to which pension expansions reach economically vulnerable populations and the possibility that non-labor income may alter work incentives or household investment decisions \citep{bando2022, avila2024}.

This paper evaluates the most ambitious expansion of Mexico’s Social Pension Program (PAM) to date. Beginning in 2019, the reform simultaneously increased benefit generosity and expanded eligibility to individuals previously excluded because they received contributory pensions. The reform marked the first stage of a broader transition toward universal coverage, which was completed in 2022. As a result, the Mexican experience provides a unique opportunity to examine a central question in the design of social protection systems: what are the consequences of replacing a targeted pension scheme with a universal one?

We make three main contributions. First, we provide new causal evidence on the effects of universalizing a non-contributory pension program in a middle-income country. While most existing studies focus on targeted programs or incremental expansions in coverage, considerably less is known about the consequences of extending benefits to previously ineligible groups, particularly individuals who already receive contributory pensions. Second, we separately identify the effects of expanded eligibility and increased benefit generosity. Because these two policy changes occurred simultaneously, conventional evaluations cannot distinguish whether observed effects are driven by larger transfers or by broader coverage. Our triple-differences strategy exploits variation across time, age, and pension status to isolate the effect of universalization from that of the transfer increase. Third, we show that the effects of pension reforms extend beyond direct beneficiaries. By examining multigenerational households, we document spillover effects on consumption patterns, health investments, educational expenditures, and school attendance among younger household members. These findings contribute to a growing literature showing that social protection programs can influence human capital accumulation through household-level reallocations of resources, labor supply adjustments, and intergenerational transfers.

Our findings indicate that the two dimensions of the reform had markedly different consequences. The combined reform significantly increased program participation, raised household income, and reduced both poverty and extreme poverty. However, these gains were driven primarily by the increase in transfer generosity. Once we isolate the universalization component, we find no evidence of poverty reduction and, if anything, a modest increase in extreme poverty by 2024. Consistent with the predictions of a standard labor-leisure framework, this increase is concentrated among low-income elderly individuals who appear to have reduced labor supply following enrollment in the program, despite the transfer being insufficient to fully replace forgone earnings.

The analysis further reveals substantial heterogeneity and important household spillovers. Universalization expanded participation among low-income elderly individuals who may previously have faced barriers to enrollment, while also generating substantial take-up among higher-income beneficiaries whose economic behavior remained largely unchanged. Moreover, the reform affected household consumption patterns, health investments, and educational expenditures differently across the income distribution, suggesting that the effects of universal social pensions extend beyond direct beneficiaries and may influence broader household decisions related to human capital accumulation.

Taken together, our findings highlight an important trade-off in the design of social pensions. Universalization can reduce exclusion and simplify access, particularly for disadvantaged populations, but it may also broaden the distribution of public resources beyond economically vulnerable groups and generate behavioral responses that partially offset some of the intended welfare gains. More broadly, the results underscore the importance of distinguishing between the effects of expanding coverage and increasing benefit generosity when evaluating social protection programs. This distinction is likely to become increasingly relevant as middle-income countries confront the challenges of population aging while seeking to maintain equitable and fiscally sustainable systems of old-age income support.

\section{Policy background}
\label{Background}
Mexico's retirement system combines contributory and non-contributory components. Contributory pensions are primarily provided through social security institutions such as the Instituto Mexicano del Seguro Social (IMSS) and the Instituto de Seguridad y Servicios Sociales de los Trabajadores del Estado (ISSSTE), which cover private- and public-sector workers, respectively. Eligibility for these pensions depends on an individual's contribution history and participation in formal employment. Given Mexico's historically high levels of labor informality, a substantial share of older adults reach retirement age without qualifying for a contributory pension, creating an important gap in old-age income protection.

\begin{figure}[ht!]
\centering
\caption{Evolution of PAM in Mexico}
  \centering
  \includegraphics[scale=0.65]{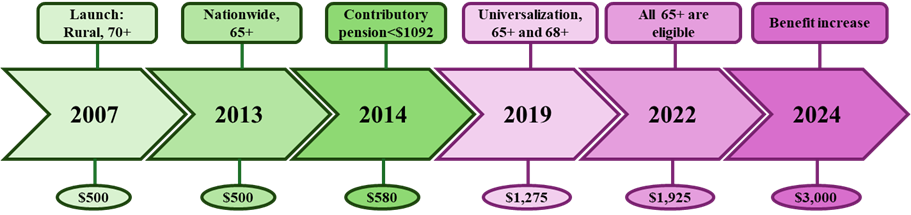}
\hfill
\fnote{\footnotesize \textit{Note:} The social pension program (PAM) was launched in Mexico in 2007. PAM grew over time in terms of eligibility and the size of the cash transfer. PAM became universal in 2019.}
\label{figPolicyBack}
\end{figure}

Mexico’s non-contributory pension program, hereafter referred to as the Mexican Social Pension Program (PAM, for its acronym in Spanish), was introduced to address this coverage gap among older adults excluded from the contributory pension system. The program was initially designed to improve income security for older adults in rural areas, targeting individuals aged 70 and older residing in localities with fewer than 2,500 inhabitants. Over time, eligibility expanded geographically and demographically, with national coverage reached in 2012. In 2013, the eligibility age was reduced to 65, although restrictions remained for individuals receiving contributory pensions or with incomes above a modest threshold. The monthly transfer amount also remained relatively modest during this period, increasing only gradually from 500 to 580 pesos (approximately 43 USD; see \autoref{figPolicyBack} for a graphical summary of the program's evolution).

A major policy shift occurred in 2019 under the administration of President Andrés Manuel López Obrador. The reform consisted of two key elements. First, the monthly benefit increased substantially, rising from 580 to 1,275 pesos (approximately 66 USD). Further increases followed in subsequent years, with the benefit reaching 3,000 pesos (approximately 176 USD) by 2024. Second, the reform initiated a phased expansion in eligibility that ultimately led to universal coverage. Initially, Indigenous individuals aged 65 and older and non-Indigenous individuals aged 68 and older became eligible regardless of their contributory pension status, with priority given to marginalized and rural communities. The expansion culminated in 2022, when the program became universal for all individuals aged 65 and older, regardless of income, contributory pension status, or place of residence.

In this paper, we focus on the initial phase of this transformation, namely the 2019 expansion, which provides a unique opportunity to examine the effects of simultaneously increasing benefit generosity and expanding program coverage. Our identification strategy exploits this sharp policy change to estimate the causal effects of the reform on key socioeconomic outcomes.

It is worth noting that the 2019 reform coincided with other major policy developments, including the termination of the PROSPERA conditional cash transfer program and substantial increases in the minimum wage. We address the potential confounding effects of these concurrent reforms in dedicated robustness checks presented later in the paper.

\section{Theoretical framework}
\label{TFramework}
The main empirical finding of this paper is that the universalization of the Mexican social pension program (PAM) increased extreme poverty among some low-income elderly individuals by reducing their participation in the labor market. To interpret this result, we develop a simple consumption-leisure framework in which the pension operates as a source of non-labor income. The framework illustrates how the introduction of a universal pension may generate heterogeneous labor supply responses across income groups and provides testable predictions for the empirical analysis.

Consider an elderly individual who derives utility from consumption ($c$) and leisure ($l$). The utility function $u(c,l)$ satisfies the standard assumptions:
\[
MU_{c}=\frac{\partial \thinspace u(c,l)}{\partial \thinspace c}>0 \thinspace\thinspace ; \thinspace\thinspace\thinspace\thinspace MU_{l}=\frac{\partial \thinspace u(c,l)}{\partial \thinspace l}>0
\]
\[
\frac{\partial^2 \thinspace u(c,l)}{\partial \thinspace c^2}<0 \thinspace\thinspace ; \thinspace\thinspace\thinspace\thinspace \frac{\partial^2 \thinspace u(c,l)}{\partial \thinspace l^2}<0
\]
The individual allocates a fixed unit of time between labor ($L$) and leisure ($l$), such that:
\[
L=1-l
\]
Consumption goods have a price $p$. Income is obtained from labor earnings and from a government transfer. Let $w$ denote the wage rate and $m$ the non-contributory pension received through PAM. The individual's optimization problem is:
\[
\max_{c,l} \thinspace\thinspace\thinspace u(c,l)
\]
subject to
\[
p \thinspace c=m+w \thinspace L
\]
Substituting the time constraint yields:
\[
p \thinspace c=m+w \thinspace (1-l)
\]
The optimal allocation satisfies the standard condition that the marginal rate of substitution between leisure and consumption equals the real wage:
\[
MRS_{l,c}=\frac{w}{p}
\]
Graphically, the optimum corresponds to the tangency between the budget constraint and the individual's indifference curve.

\begin{figure}[ht!]
\centering
\caption{Consumption-leisure problem for young and elderly individuals}
\begin{subfigure}{.49\textwidth}
  \centering
  \includegraphics[width=1\linewidth]{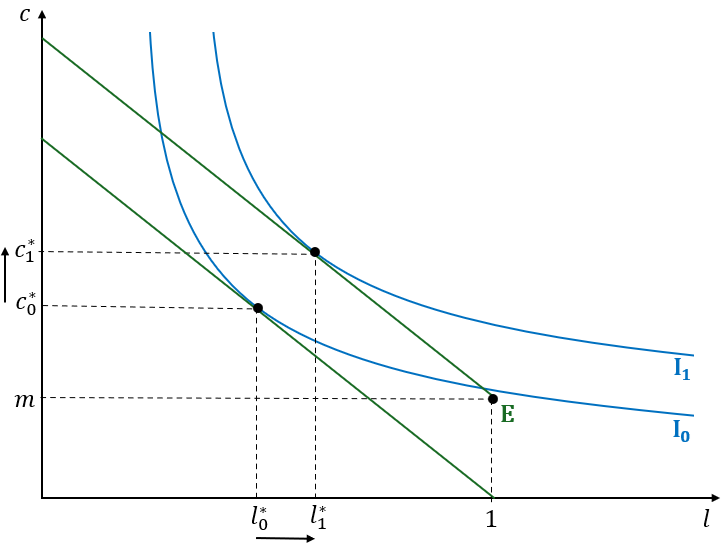}
  \caption{Elderly optimal choice}
\end{subfigure}
\hfill
\begin{subfigure}{.49\textwidth}
  \centering
  \includegraphics[width=1\linewidth]{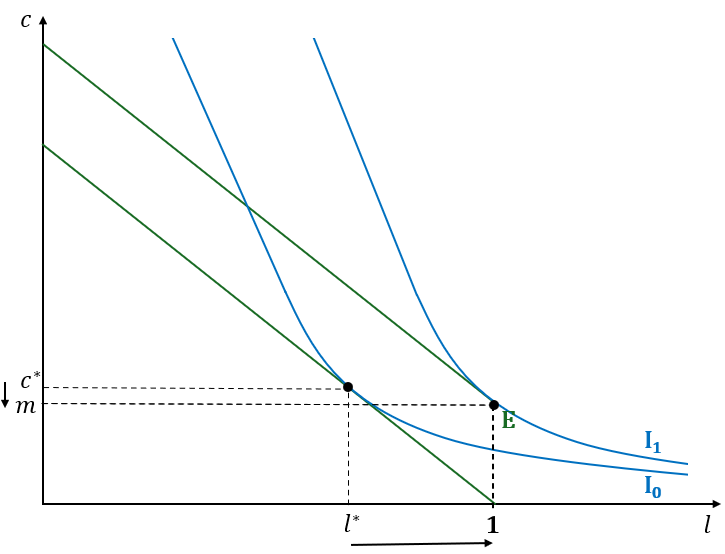}
  \caption{Optimal choice with skewed preferences}
\end{subfigure}
\hfill
\fnote{\footnotesize \textit{Note:} Panel (a) represents the optimal choice for the standard consumption-leisure model, with and without the PAM cash transfer. Panel (b) does the same assuming skewed preferences to represent a case where elderly value leisure more than consumption goods.}
\label{TheoFram}
\end{figure}

The introduction of PAM increases non-labor income and shifts the budget constraint outward by the amount of the transfer, $m$. If leisure is a normal good, the transfer generates a positive income effect that increases desired leisure and reduces labor supply. For a representative individual, the optimal allocation moves from $(l_0^{},c_0^{})$ in the absence of the transfer ($m=0$) to $(l_1^{},c_1^{})$ when the social pension is introduced ($m>0$). This outcome is illustrated in Panel (a) of Figure~\ref{TheoFram}.

Importantly, the magnitude of this response is expected to vary across individuals. For high-income elderly, the pension represents a relatively small share of total income and therefore generates only a modest change in the budget set. In contrast, for low-income elderly, the pension may constitute a substantial fraction of pre-transfer income, generating a larger income effect and a stronger incentive to substitute leisure for labor. As a result, reductions in labor supply should be concentrated among individuals with lower initial income levels.

Panel (b) of Figure~\ref{TheoFram} illustrates an extreme case in which the transfer is sufficiently valuable relative to labor earnings that the individual chooses full retirement. In this situation, the pension acts as a substitute for labor income and the elderly prefers to consume more leisure despite a reduction in consumption relative to the pre-transfer allocation. More generally, however, the theoretical framework does not require complete retirement. Depending on the magnitude of the transfer and the individual's preferences, the introduction of PAM may lead to partial reductions in labor supply, no change in employment, or complete withdrawal from the labor market.

The framework generates two empirical predictions that guide our analysis. First, the introduction of a non-contributory pension may reduce labor supply through a positive income effect if leisure is a normal good. Second, this response should be strongest among low-income elderly, for whom the transfer represents a larger share of total resources. The effect on poverty is theoretically ambiguous because it depends on the balance between the additional transfer income and any reduction in labor earnings induced by the pension. These predictions are consistent with the patterns documented in the empirical analysis.

\section{Data}
\label{Data}
We use data from the Socioeconomic Conditions Module (MCS) of Mexico’s National Household Income and Expenditure Survey (ENIGH) to conduct our analysis. Specifically, we pool the 2016, 2018, 2020, 2022, and 2024 rounds to construct a repeated cross-sectional dataset. The ENIGH is a nationally representative survey administered by INEGI that collects detailed demographic and economic information from Mexican households, with representativeness at the national, state, and urban–rural levels. It reports comprehensive data on each household member’s income sources, including labor earnings, contributory pensions, retirement payments, and non-contributory transfers such as those from the PAM program.

Our empirical analysis focuses on four key outcomes: (i) PAM participation, (ii) (extreme) poverty status, (iii) income, and (iv) employment. Participation in PAM is defined as a binary indicator equal to one if the individual reports receiving a transfer from the program. Income is measured as total monthly household income per adult equivalent, which includes both monetary income (from wages, independent work, capital income, and transfers) and non-monetary income (such as in-kind payments, gifts, or self-consumption), divided by the number of adult-equivalent household members using the standard equivalence scale employed by CONEVAL.

To construct poverty measures, we follow Mexico’s official methodology \citep{coneval2014}. The extreme poverty line reflects the cost of a basic food basket that satisfies the minimum daily caloric intake for the average Mexican, while the moderate poverty line adds a non-food basket covering essential items such as hygiene products, clothing, and public transportation. Both thresholds are estimated separately for urban and rural areas, accounting for cost-of-living differences \citep{licona2016}. We define binary indicators for poverty and extreme poverty equal to one if the individual’s per capita income falls below the corresponding threshold.

Employment is captured using a binary variable equal to one if the respondent reported working in the month prior to the survey, based on the question: “Did you work last month?” Note that we are able to measure both formal and informal employment. However, we cannot distinguish between them given the information provided. We also examine occupational composition by defining a self-employment indicator equal to one if the individual reports working in their own business or enterprise.\footnote{Based on the question: “Did you work in your own business or venture?”}

To improve the precision of our estimates, we include a standard set of control variables: sex, years of formal education, self-reported indigenous status, and cohabitation status (equal to one if the elderly person lives with non-elderly household members). We also control for locality size using four categories: fewer than 2,500 inhabitants; 2,500–14,999; 15,000–99,999; and more than 100,000. These categories help account for variation in local labor market conditions and access to social infrastructure.

In addition, all regressions include survey year fixed effects to absorb national-level shocks (such as the COVID-19 pandemic) and state-specific linear trends to account for time-varying unobserved heterogeneity at the state level, such as economic conditions or the staggered implementation of complementary social programs.

To contextualize our empirical strategy, we present summary statistics for key variables before and after the 2019 PAM expansion, disaggregated by treatment and control groups and further stratified by contributory pension status. The treatment group comprises individuals aged 68–70, while the control group includes those aged 62–64. Within each age group, we distinguish between individuals with and without a contributory pension. \autoref{ss} shows that, once we account for this triple interaction, we do not observe systematic differences across groups in baseline characteristics, supporting the validity of our identification strategy. As expected, PAM take-up among the control group is nearly zero in all years, reinforcing the credibility of our comparison. Additionally, poverty and employment rates are substantially higher among individuals without access to a contributory pension (nearly twice as high as those with a pension), highlighting the economic vulnerability of the population targeted by the universalization reform. These patterns underscore the importance of disaggregating effects by pension status to accurately capture the policy's causal impact.

\section{Empirical Strategy}
\label{EmpiricalS}
To estimate the causal effect of the universalization of Mexico’s non-contributory pension program, we use the exogenous variation produced by the 2019 PAM universalization. In particular, we exploit variation along three dimensions: age-based eligibility, time (pre- and post-reform), and pension scheme status. This triple-differences (DDD) framework allows us to isolate the impact of expanded eligibility from contemporaneous increases in the benefit amount, which would otherwise confound standard difference-in-differences (DiD) estimates. Our approach mirrors the empirical strategy of \citet{gruber1994}, who used a DDD design to disentangle the effects of mandated maternity benefits from other labor market trends in the United States.

Importantly, the DDD strategy helps mitigate some common concerns associated with DiD designs by controlling for additional sources of heterogeneity across treatment and comparison groups. Nevertheless, we assess the credibility of the identifying assumptions underlying both our DiD and DDD specifications. In the next section, we present supporting evidence of parallel pre-trends for our DiD estimates. Similarly, following the approach recommended by \citet{olden2022}, we examine the plausibility of the DDD identification assumption by testing for differential pre-trends across the triple interaction components.

A baseline DiD specification compares individuals just above and just below the age eligibility threshold, before and after the reform. This reduced-form design captures the combined effect of universalization and the increase in the cash transfer amount:
\begin{equation}
y_{iat} = \alpha + \beta \, (Treat_a \times After_t) + \mu_a + \mu_t + \mu_{st} + \boldsymbol{X}_{iat}\lambda + \varepsilon_{iat}
\label{did_eq}
\end{equation}

where $y_{iat}$ is the outcome of interest for individual $i$ of age $a$ at time $t$, $Treat_a$ is an indicator for being in the eligible age group (68–70), and $After_t$ is an indicator for the post-universalization period. The coefficient $\beta$ captures the intention-to-treat effect of the 2019 policy change, but conflates changes in eligibility with changes in transfer size.

To disentangle these effects, we introduce a third dimension of variation based on pension scheme status. Specifically, we compare eligible individuals with a contributory pension, who became newly eligible after the reform, to otherwise similar individuals without a contributory pension, who remained eligible throughout. Our estimating equation is:
\begin{equation}
\begin{aligned}
y_{ijat} = \, & \alpha + \beta \, (Treat_a \times After_t \times PS_j) 
+ \delta_1 (Treat_a \times After_t) 
+ \delta_2 (Treat_a \times PS_j) \\
& + \delta_3 (After_t \times PS_j) 
+ \delta_4 PS_j 
+ \mu_a + \mu_t + \mu_{st} + \boldsymbol{X}_{ijat}\lambda + \varepsilon_{ijat}
\end{aligned}
\label{DDD}
\end{equation}
Here, $PS_j$ is a binary indicator equal to one for individuals with a contributory pension exceeding the previous eligibility threshold (i.e., more than 1,092 pesos monthly), who became newly eligible due to the universalization. The coefficient of interest, $\beta$, captures the differential change in outcomes for this newly treated group, netting out the increase in benefit amounts, time trends, age-specific effects, and differences across pension groups unrelated to the reform.

All specifications include age fixed effects ($\mu_a$), year fixed effects ($\mu_t$), and individual-level controls ($\boldsymbol{X}_{ijat}$), including gender, years of education, indigenous status, cohabitation status, and locality size. We also include state-specific linear time trends, $\mu_{st}$, to absorb unobserved regional variation in labor markets, poverty rates, or exposure to other social programs.

Following \citet{abadie2023}, we cluster standard errors at the municipality level in all specifications. As argued by \citet[p.~33]{abadie2023}, when treatment is assigned at the cluster level within a random sample, standard errors should be clustered at the level of treatment assignment. However, when the data are drawn from a population using a clustered sampling design, standard errors should instead be clustered at the level at which sampling first occurs. Our setting falls into the latter category. The ENIGH survey is based on a multi-stage stratified sampling procedure. In the first stage, a set of primary sampling units (PSUs) is randomly selected, and in the second stage, households are sampled within these PSUs. According to Mexico’s National Institute of Statistics and Geography (INEGI), these PSUs correspond most closely to municipalities, particularly in rural areas. Therefore, clustering at the municipality level appropriately accounts for the structure of the sampling design and yields robust inference.

To assess the validity of our identifying assumption, we estimate an event-study version of our DDD model:
\begin{equation}
\begin{aligned}
y_{ijat} = \, & \alpha + \sum_{t \neq 2018} \beta_t \cdot I(treatment_{ajt} = t) 
+ \delta_1 (Treat_a \times After_t) 
+ \delta_2 (Treat_a \times PS_j) \\
& + \delta_3 (After_t \times PS_j) 
+ \delta_4 PS_j 
+ \mu_a + \mu_t + \mu_{st} + \boldsymbol{X}_{ijat}\lambda + \varepsilon_{ijat}
\end{aligned}
\label{event}
\end{equation}
In this specification, $I(treatment_{ajt} = t)$ is a set of year-specific indicators equal to one if individual $i$ belongs to the treated group (eligible and with a contributory pension) and is observed in year $t$. We normalize 2018, the last pre-reform year, as the reference category. The coefficients $\beta_t$ trace the dynamic treatment effects over time, allowing us to assess both the timing and magnitude of the policy impact, as well as the presence of any pre-treatment trends.

\section{Results}
\subsection{Universalization of social pensions and effects on poverty and labor market outcomes}
\subsubsection{Combined effects: Increase in cash transfer and universalization}
We begin by evaluating the overall impact of the 2019 reform to Mexico’s non-contributory pension program, which simultaneously expanded eligibility and increased the monthly transfer amount. To estimate the combined effect of these policy changes, we implement the standard difference-in-differences (DiD) specification introduced in \autoref{did_eq}.

\begin{figure}[ht!]
\centering
\caption{Effects of the expansion of social pensions in Mexico (coverage and cash transfer)}
\begin{subfigure}{.49\textwidth}
  \centering
  \includegraphics[width=1\linewidth]{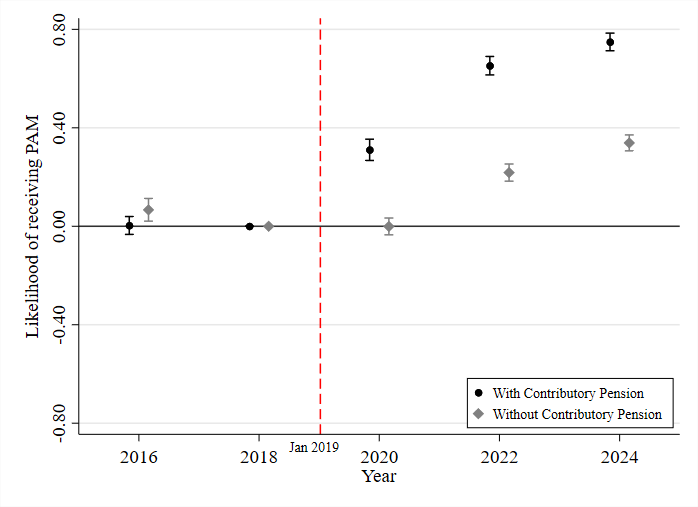}
  \caption{Program take-up}
\end{subfigure}
\hfill
\begin{subfigure}{.49\textwidth}
  \centering
  \includegraphics[width=1\linewidth]{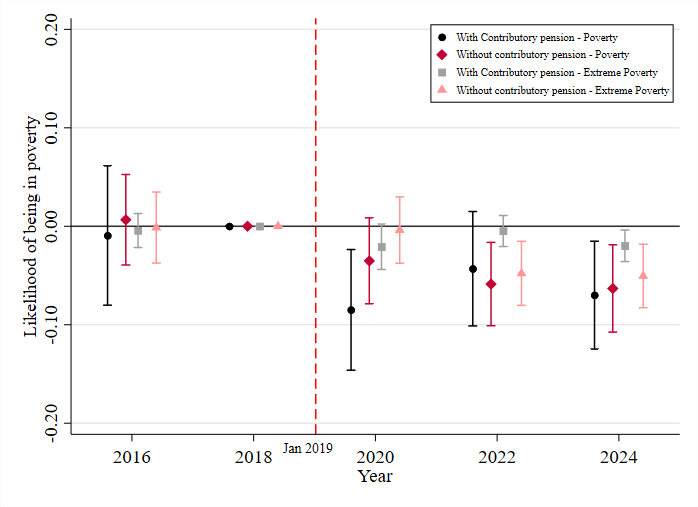}
  \caption{Poverty outcomes}
\end{subfigure}
\hfill
\begin{subfigure}{.49\textwidth}
  \centering
  \includegraphics[width=1\linewidth]{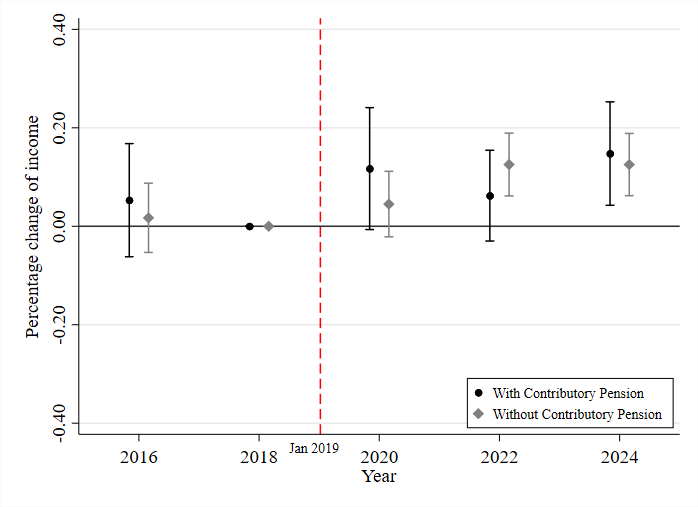}
  \caption{Ln(income)}
\end{subfigure}
\hfill
\begin{subfigure}{.49\textwidth}
  \centering
  \includegraphics[width=1\linewidth]{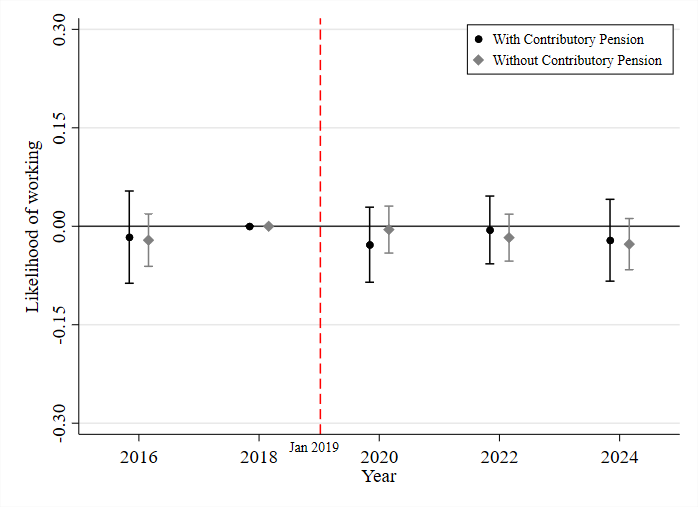}
  \caption{Work}
\end{subfigure}
\hfill
\fnote{\footnotesize \textit{Note:} Graph points show differences in outcomes between treatment and control groups relative to 2018 levels. Estimates come from the event-study version of \autoref{did_eq}. Whiskers show 95\% confidence intervals. The vertical line depicts the date when the policy intervention was enacted (January 2019). This figure provides suggestive evidence on the parallel trend assumption underlying our difference-in-differences identification strategy. Data are from the 2016, 2018, 2020, 2022, and 2024 Household Income and Expenditure National Survey (ENIGH).}
\label{figMainDiDR}
\end{figure}

Panel (a) of \autoref{figMainDiDR} documents a substantial increase in program take-up among the newly eligible population following the reform. The figure reports estimates separately for individuals with and without contributory pensions, allowing us to visualize the differential effects of the reform across groups. Because individuals without contributory pensions were already eligible for the program before the reform, while those with contributory pensions became newly eligible as a result of the universalization, the difference in outcomes between these two groups provides an initial indication of the contribution of the eligibility expansion. As we show formally in the next subsection using a triple-differences framework, most of the increase in program participation appears to be driven by the universalization component of the reform. By 2024, take-up rates among newly eligible individuals exceeded 70\%.

Panel (b) of \autoref{figMainDiDR} shows that both poverty and extreme poverty declined significantly following the reform. On average, the policy reduced overall poverty by 5.8 percentage points and extreme poverty by 2.6 percentage points (see Panel A, columns (3) and (4) of \autoref{tab2}). This pattern contrasts with the findings of \citet{avila2024}, who show that prior to 2019 the program was more effective at reducing extreme poverty than broader measures of poverty. The larger effect on poverty after 2019 suggests that the substantial increase in the transfer amount enabled households located just below the poverty line to move above it, whereas many individuals in extreme poverty may have already been covered by earlier versions of the program.

Panel (c) highlights the most likely mechanism behind these improvements: a sustained increase in household income. According to the estimates reported in column (2) of \autoref{tab2}, the reform increased per capita household income by approximately 9\%. These gains closely track the successive increases in the transfer amount and appear to be the principal channel through which the reform reduced poverty.

However, Panel (d) also reveals a modest decline in employment in 2024. Although the average effect is relatively small, the figure suggests that employment reductions are more pronounced among newly eligible individuals with contributory pensions. This pattern raises the possibility that part of the income gain generated by the transfer was offset by reductions in labor earnings. We explore this mechanism in greater detail in \autoref{Mechanisms}, where we examine heterogeneous labor supply responses across income groups.

These results indicate that the combined reform substantially increased program participation and reduced poverty, primarily through higher transfer income. At the same time, the heterogeneity across pension groups suggests that the universalization component may have generated distinct behavioral responses that are not apparent when focusing solely on average effects. The next subsection formally isolates these effects using our triple-differences identification strategy.

\subsubsection{Isolating the Effect of Universalization}
To disentangle the effects of expanded eligibility from those of increased transfer amounts, we turn to the triple-differences (DDD) framework and event-study design outlined in \autoref{EmpiricalS}. This strategy allows us to isolate the causal effect of the universalization by netting out income effects common to both newly and previously eligible individuals.

\begin{figure}[ht!]
\centering
\caption{Effects of the universalization of social pensions in Mexico}
\begin{subfigure}{.49\textwidth}
  \centering
  \includegraphics[width=1\linewidth]{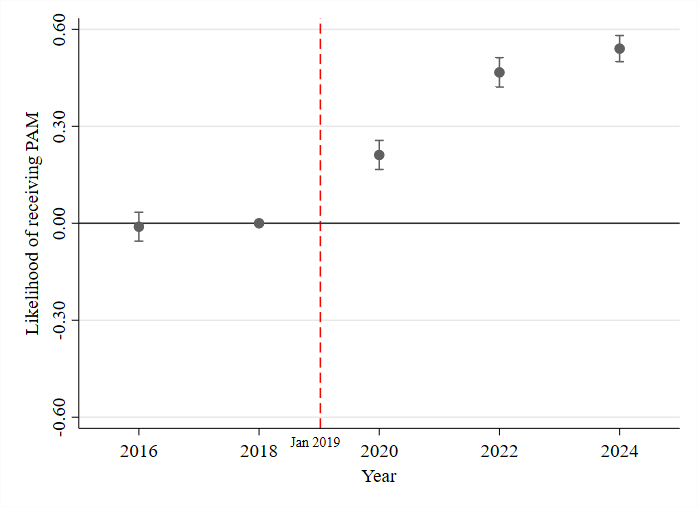}
  \caption{Program take-up}
\end{subfigure}
\hfill
\begin{subfigure}{.49\textwidth}
  \centering
  \includegraphics[width=1\linewidth]{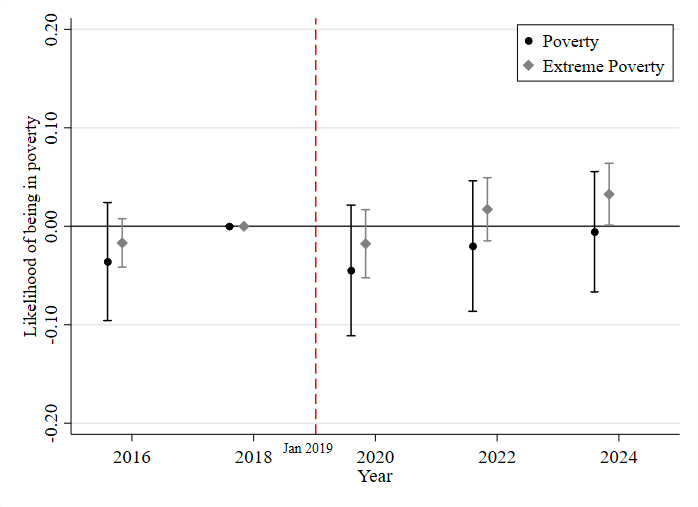}
  \caption{Poverty outcomes}
\end{subfigure}
\hfill
\begin{subfigure}{.49\textwidth}
  \centering
  \includegraphics[width=1\linewidth]{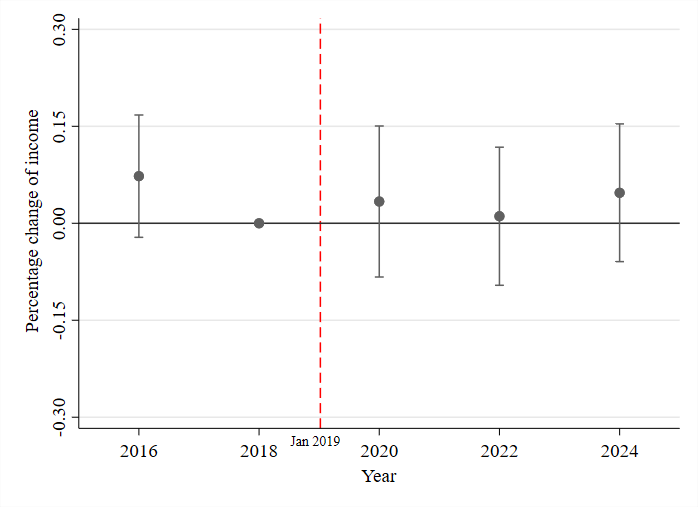}
  \caption{Ln(income)}
\end{subfigure}
\hfill
\begin{subfigure}{.49\textwidth}
  \centering
  \includegraphics[width=1\linewidth]{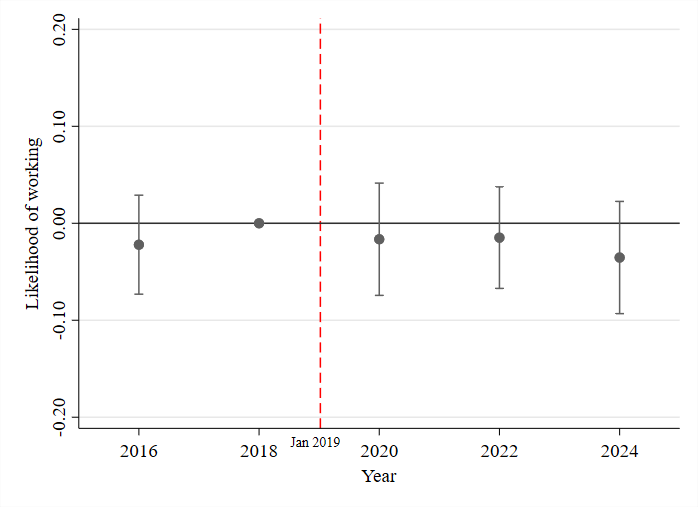}
  \caption{Work}
\end{subfigure}
\hfill
\fnote{\footnotesize \textit{Note:} Graph points show differences in outcomes between treatment and control groups relative to 2018 levels, between new eligible individuals with a contributory pension and the always eligible without a contributory pension. Estimates come from \autoref{event}. Whiskers show 95\% confidence intervals. The vertical line depicts the date when the policy intervention was enacted (January 2019). This figure provides suggestive evidence on the parallel trend assumption underlying our DDD identification strategy. Data are from the 2016, 2018, 2020, 2022, and 2024 Household Income and Expenditure National Survey (ENIGH).}
\label{figMainR}
\end{figure}

Panel (a) of \autoref{figMainR} confirms that the sharp post-reform increase in participation is largely attributable to the expansion in eligibility. The newly eligible group, those with contributory pensions exceeding the prior exclusion threshold, exhibited a marked increase in take-up following the reform.

However, Panel (b) reveals a more nuanced outcome. The universalization itself appears to have had no statistically significant effect on poverty. If anything, extreme poverty rose slightly in 2024, though the magnitude is modest. This stands in contrast to the DiD estimates and suggests that the marginal beneficiaries of the reform were relatively better off and therefore less likely to experience large welfare gains. By extending coverage to this group, the program’s targeting efficiency may have declined, diluting its aggregate poverty-reducing effect.

Panels (c) and (d) support this interpretation. Among the newly eligible, we find no discernible changes in household income or employment outcomes. These null results imply that the additional pension income did not meaningfully alter economic behavior or living standards among the higher-income elderly population.

The modest rise in extreme poverty among the treated group in 2024 warrants further examination. One potential explanation is that the expansion triggered labor market adjustments or discouraged labor force participation among economically vulnerable individuals. We explore this hypothesis in greater detail in \autoref{Mechanisms}.

\subsection{Mechanisms}
\label{Mechanisms}
\subsubsection{Income effect and the opportunity cost of leisure}
The theoretical framework developed in Section~\ref{TFramework} predicts that the universalization of the pension program may affect labor supply through a standard income effect. By increasing non-labor income, the pension raises the demand for leisure if leisure is a normal good. Moreover, the framework suggests that this response should be strongest among low-income elderly individuals, for whom the transfer represents a larger share of total resources. The ultimate effect on poverty is therefore theoretically ambiguous and depends on the balance between the additional transfer income and any reduction in labor earnings induced by the program.

To evaluate these predictions, we estimate heterogeneous effects of the reform across the income distribution. Specifically, we examine the impact of the universalization on program take-up and employment decisions by income quartile, allowing us to identify differential behavioral responses among low- and high-income elderly individuals.

\begin{figure}[ht!]
\centering
\caption{Mechanism: Income effect}
\begin{subfigure}{.49\textwidth}
  \centering
  \includegraphics[width=1\linewidth]{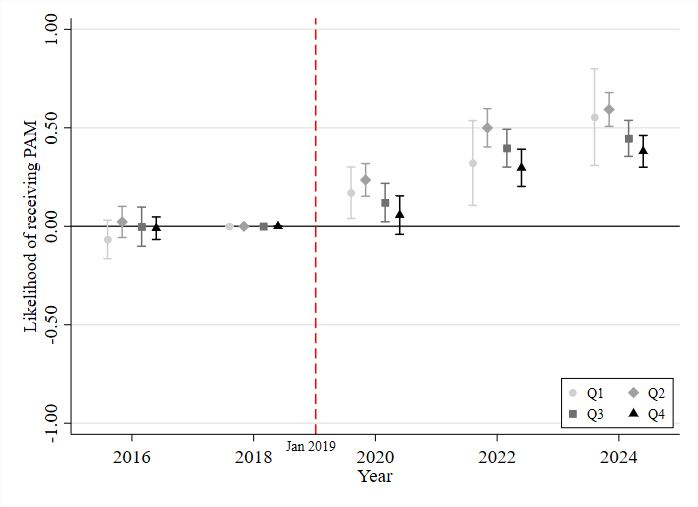}
  \caption{Program take-up by quartiles of income}
\end{subfigure}
\hfill
\begin{subfigure}{.49\textwidth}
  \centering
  \includegraphics[width=1\linewidth]{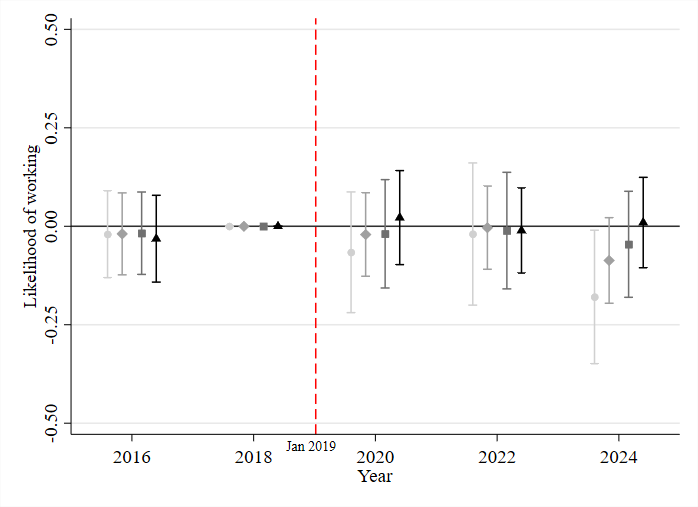}
  \caption{Work by quartiles of income}
\end{subfigure}
\hfill
\begin{subfigure}{.49\textwidth}
  \centering
  \includegraphics[width=1\linewidth]{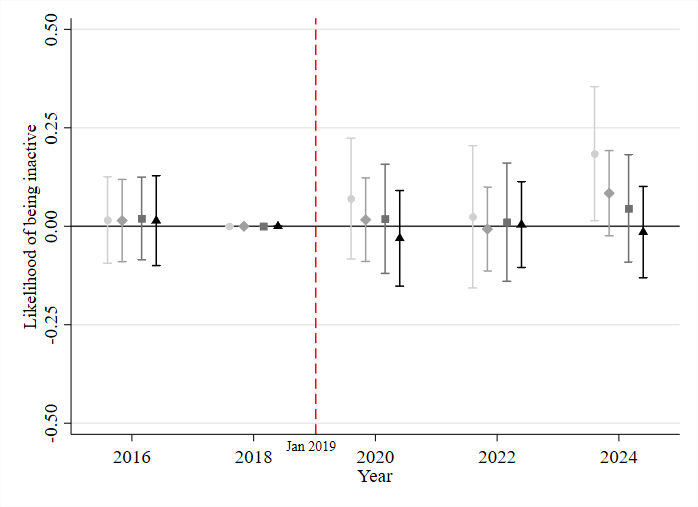}
  \caption{Inactive by quartiles of income}
\end{subfigure}
\hfill
\begin{subfigure}{.49\textwidth}
  \centering
  \includegraphics[width=1\linewidth]{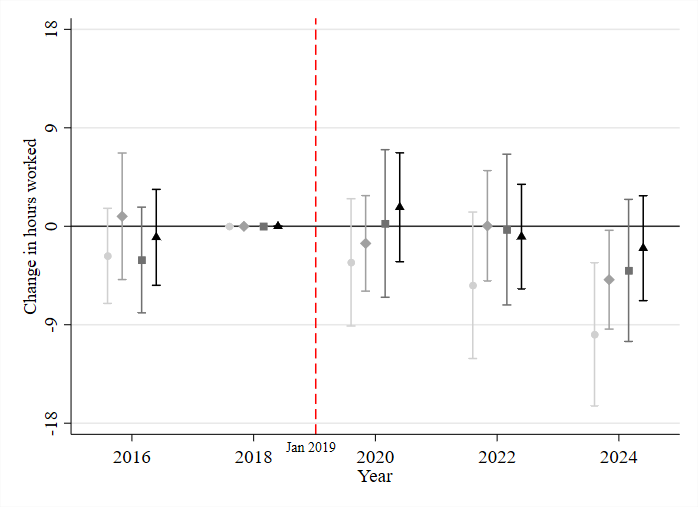}
  \caption{Hours worked by quartiles of income}
\end{subfigure}
\hfill
\fnote{\footnotesize \textit{Note:} Graph points show differences in outcomes between treatment and control groups relative to 2018 levels, between new eligible individuals with a contributory pension and the always eligible without a contributory pension. Estimates come from \autoref{event}. Each color corresponds to separate regressions conditioning on income quartiles. Whiskers show 95\% confidence intervals. The vertical line depicts the date when the policy intervention was enacted (January 2019). This figure provides suggestive evidence on the parallel trend assumption underlying our DDD identification strategy. Data are from the 2016, 2018, 2020, 2022, and 2024 Household Income and Expenditure National Survey (ENIGH).}
\label{figMechanismIncome}
\end{figure}

Panel (a) of \autoref{figMechanismIncome} shows that take-up rates increased throughout the income distribution, although the magnitude of the increase varied substantially. The second income quartile experienced the largest increase in participation, while the bottom quartile, composed largely of individuals already covered under the pre-2019 scheme, exhibited more modest gains.

Importantly, however, we also observe a meaningful increase in participation among the lowest-income elderly. This finding suggests that the universalization succeeded in expanding access among economically vulnerable individuals who may have previously faced administrative, informational, or documentation barriers to enrollment \citep{Finkelstein2019,Bhargava2015}. By simplifying eligibility and eliminating the contributory pension screening requirement, the reform appears to have reduced exclusion among a segment of the population that the program was originally intended to serve.

At the upper end of the income distribution, take-up rates decline monotonically with income, as expected. Nevertheless, participation remains substantial even among the highest-income elderly, with nearly 40\% enrolling in the program after the reform. While this outcome is consistent with the objectives of a universal program, it implies that a significant share of public resources is allocated to individuals who are unlikely to experience meaningful improvements in economic well-being as a result of the transfer.

Panels (b), (c), and (d) of \autoref{figMechanismIncome} provide strong support for the predictions of the theoretical framework. The only group that exhibits a statistically significant decline in employment and an increase in inactivity following the reform is the lowest income quartile. In contrast, employment responses among middle- and high-income elderly are small and statistically indistinguishable from zero. This pattern is precisely what one would expect if the pension generates an income effect that is strongest when the transfer constitutes a large share of total household resources.

The implications for poverty are equally consistent with the framework. Although the pension increased non-labor income, the reduction in employment among low-income elderly implies a corresponding decline in labor earnings. For a subset of vulnerable individuals, the transfer was insufficient to fully compensate for lost income from work. As a result, some elderly individuals appear to have substituted labor for leisure despite becoming economically worse off, helping to explain the increase in extreme poverty observed in 2024.

These results suggest that the universalization of the pension program generated two distinct effects. On the one hand, it broadened access among disadvantaged elderly individuals who had previously remained outside the program. On the other hand, it induced labor supply reductions among the poorest beneficiaries, illustrating how expansions in non-labor income can generate unintended behavioral responses. Meanwhile, higher-income individuals largely responded by enrolling in the program without altering their labor market behavior, limiting the extent to which the additional public expenditure translated into improvements in economic welfare.

\subsubsection{Self-Employment Decisions}
Previous studies have shown that non-contributory pensions in Mexico encouraged informal self-employment among older adults. For instance, \citet{Galiani2016} and \citet{avila2024} find that, prior to the 2019 reform, recipients often used the transfer to finance small family businesses, enabling shifts into more flexible or less physically demanding work.

\begin{figure}[ht!]
\centering
\caption{Mechanism: Effects on self-employment}
\begin{subfigure}{.49\textwidth}
  \centering
  \includegraphics[width=1\linewidth]{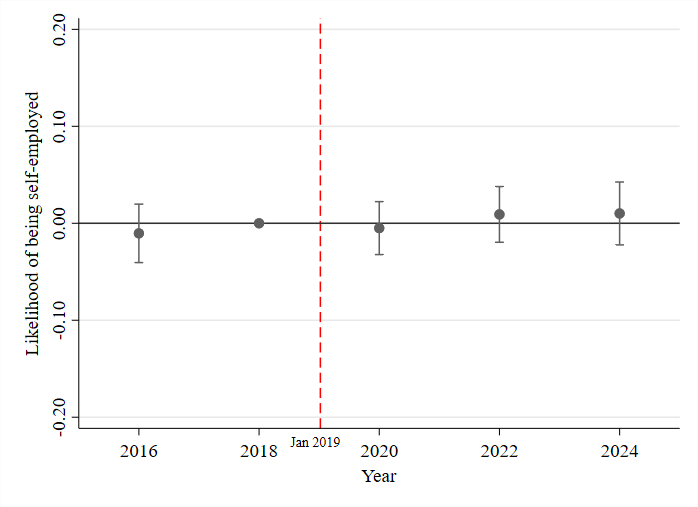}
  \caption{Combined effects}
\end{subfigure}
\hfill
\begin{subfigure}{.49\textwidth}
  \centering
  \includegraphics[width=1\linewidth]{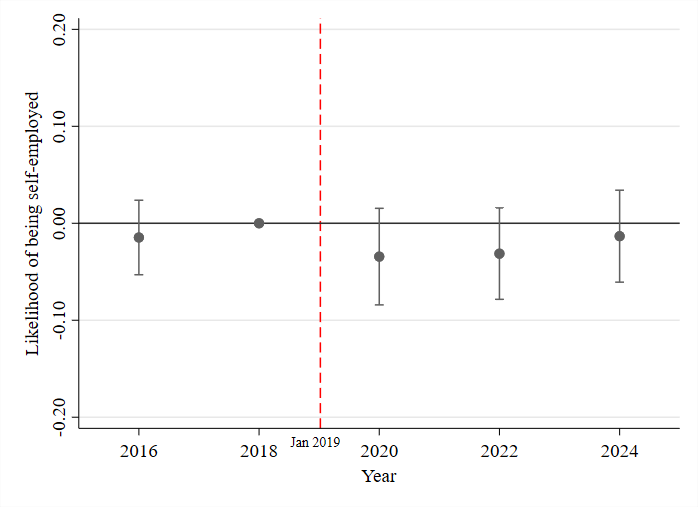}
  \caption{Universalization effect only}
\end{subfigure}
\hfill
\fnote{\footnotesize \textit{Note:} Graph points show differences in outcomes between treatment and control groups relative to 2018 levels (Panel (a), which estimates are obtained from the event-study version of \autoref{did_eq}) and also adding the difference between new eligible individuals with a contributory pension and the always eligible without a contributory pension (Panel (b), which estimates are obtained from \autoref{event}). Whiskers show 95\% confidence intervals. The vertical line depicts the date when the policy intervention was enacted (January 2019). This figure provides suggestive evidence on the parallel trend assumption underlying our DiD and DDD identification strategies. Data are from the 2016, 2018, 2020, 2022, and 2024 Household Income and Expenditure National Survey (ENIGH).}
\label{figSelfEmp}
\end{figure}

We revisit this hypothesis using both DiD and DDD specifications. Panel (a) of \autoref{figSelfEmp} presents the DiD results and shows no statistically significant change in self-employment rates following the 2019 reform. This is somewhat surprising given the larger transfer amount. One possible interpretation is that the pool of elderly willing or able to engage in informal entrepreneurship had already been exhausted during the earlier roll-out of the program.

Panel (b) isolates the effect of universalization using the DDD strategy and again finds no statistically significant effect. The point estimates are negative, suggesting that newly eligible individuals may have responded by withdrawing from the labor market altogether rather than shifting into self-employment. This is consistent with earlier results indicating an overall reduction in work among low-income groups, with no corresponding increase in informal or entrepreneurial activity.

Overall, while earlier phases of PAM may have facilitated productive reallocation within the informal sector, the universal program appears to encourage retirement or disengagement from work among newly eligible individuals.

\subsubsection{Wealth Effects}
To examine whether the universalization induced changes in household savings or asset accumulation, we constructed a wealth index using household assets and dwelling characteristics. Following \citet{galvez2018}, we apply principal component analysis (PCA) to indicators such as car ownership, housing materials, and durable goods (e.g., televisions and trucks). For each ENIGH round, we retained the minimum number of components necessary to explain at least 50\% of the variance, typically six components per year.

\begin{figure}[ht!]
\centering
\caption{Mechanism: Wealth effect}
  \centering
  \includegraphics[scale=0.5]{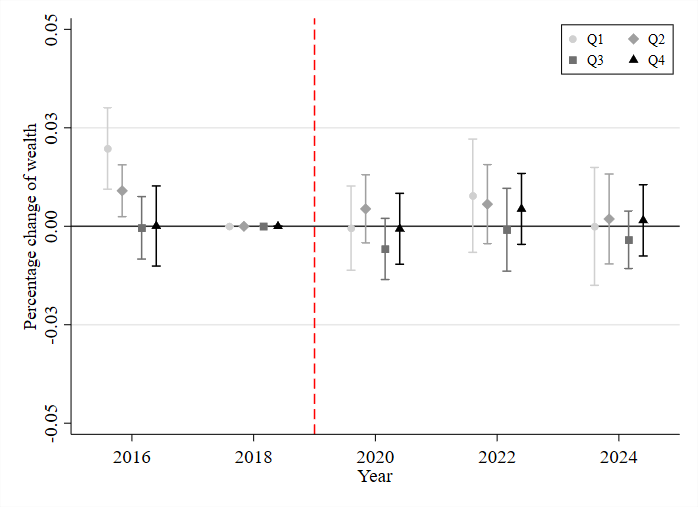}
\hfill
\fnote{\footnotesize \textit{Note:} Graph points show differences in outcomes between treatment and control groups relative to 2018 levels, between new eligible individuals with a contributory pension and the always eligible without a contributory pension. Estimates come from \autoref{event}. Each color corresponds to separate regressions conditioning on income quartiles. Whiskers show 95\% confidence intervals. The vertical line depicts the date when the policy intervention was enacted (January 2019). This figure provides suggestive evidence on the parallel trend assumption underlying our DDD identification strategy only for the two top quartiles of income distribution. Data are from the 2016, 2018, 2020, 2022, and 2024 Household Income and Expenditure National Survey (ENIGH).}
\label{Wealth}
\end{figure}

Using this index as the outcome variable, we estimate our main specifications across the full sample and by income quartile. The results, presented in \autoref{Wealth}, indicate no statistically significant effects of the reform on wealth accumulation for any income group.

This null result suggests that beneficiaries, particularly those in the upper income quartiles, did not reallocate the transfer into durable goods or long-term savings. Instead, the additional income appears to have been absorbed into regular consumption, likely non-durables. This interpretation is consistent with our earlier findings: for higher-income individuals, the transfer was too small to alter saving behavior in a meaningful way. This evidence indicates that the financial impact of PAM’s universalization was concentrated among the lower-income elderly, with minimal behavioral or welfare effects among higher-income recipients.

\subsubsection{Effects on human capital development}
\label{HC_Development}
The heterogeneity results presented in Section~\ref{Heterogeneity_Cohab} suggest that elderly individuals living in cohabitation respond differently to the universalization of the pension program than those living alone. One possible explanation is that the pension affects not only the beneficiary but also the allocation of resources within the household. In multigenerational households, pension income may be shared among family members, altering consumption choices, investments in health and education, and labor supply decisions of younger relatives. These household adjustments may, in turn, influence the incentives of elderly individuals to remain in the labor force, helping explain why labor market responses differ across living arrangements.

This mechanism is consistent with a broader literature showing that social pensions can generate spillover effects beyond direct beneficiaries. By relaxing household budget constraints, pension income may affect investments in nutrition, education, and health care, thereby influencing the accumulation of human capital among children and young adults \citep{Duflo2003, edmonds2006, Case2007}. Motivated by this literature, we focus exclusively on households in which elderly individuals live in cohabitation with younger household members and examine whether the universalization of the pension program affected consumption patterns, health investments, educational outcomes, and labor market behavior among younger generations.

\begin{figure}[ht!]
\centering
\caption{Mechanism: Changes in consumption patterns}
\begin{subfigure}{.49\textwidth}
  \centering
  \includegraphics[width=1\linewidth]{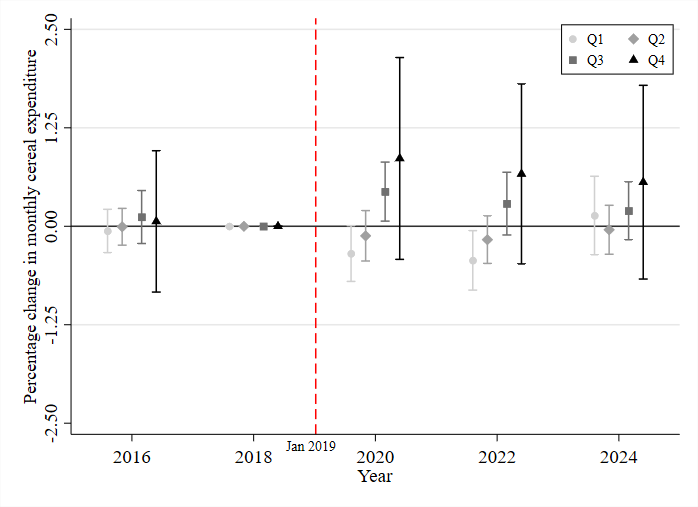}
  \caption{Cereals}
\end{subfigure}
\hfill
\begin{subfigure}{.49\textwidth}
  \centering
  \includegraphics[width=1\linewidth]{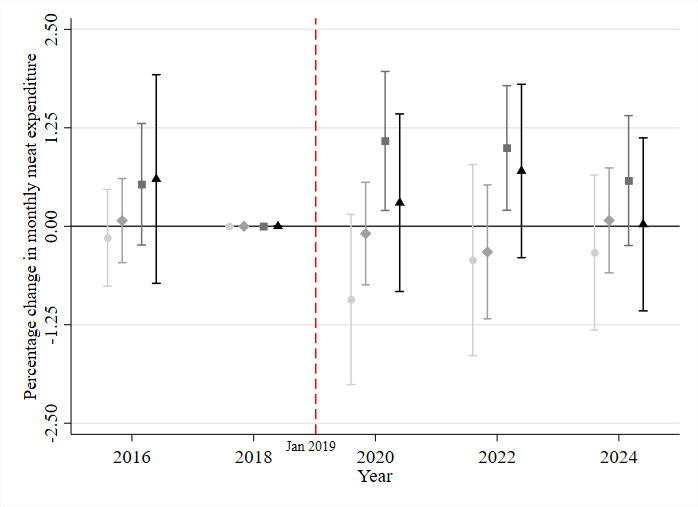}
  \caption{Meat}
\end{subfigure}
\hfill
\begin{subfigure}{.49\textwidth}
  \centering
  \includegraphics[width=1\linewidth]{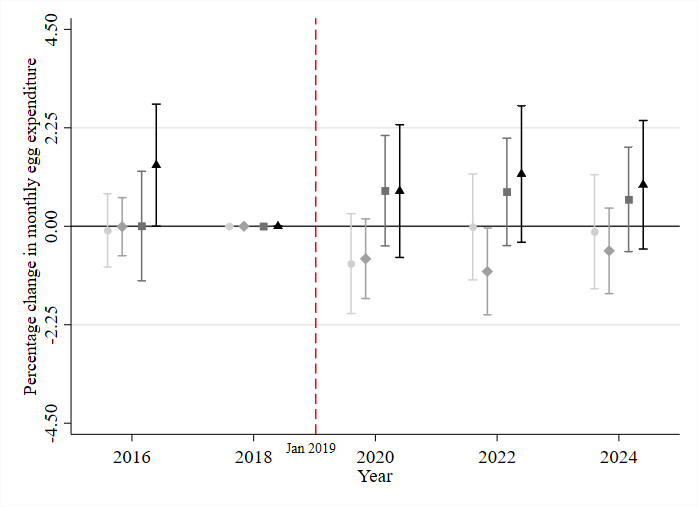}
  \caption{Eggs}
\end{subfigure}
\hfill
\begin{subfigure}{.49\textwidth}
  \centering
  \includegraphics[width=1\linewidth]{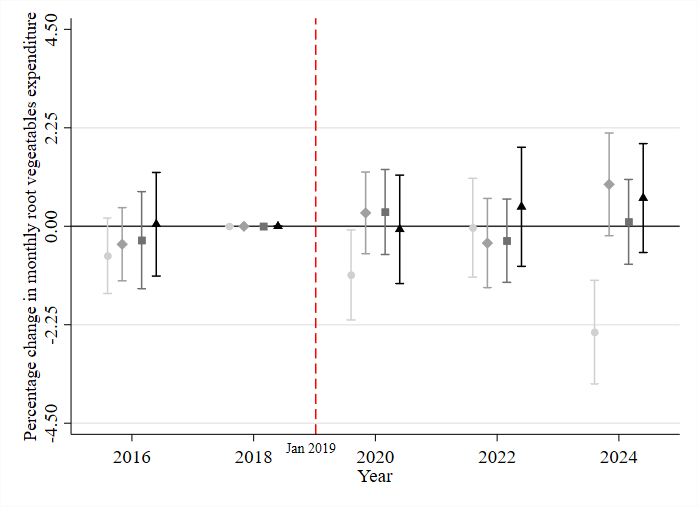}
  \caption{Root vegetables}
\end{subfigure}
\hfill
\begin{subfigure}{.49\textwidth}
  \centering
  \includegraphics[width=1\linewidth]{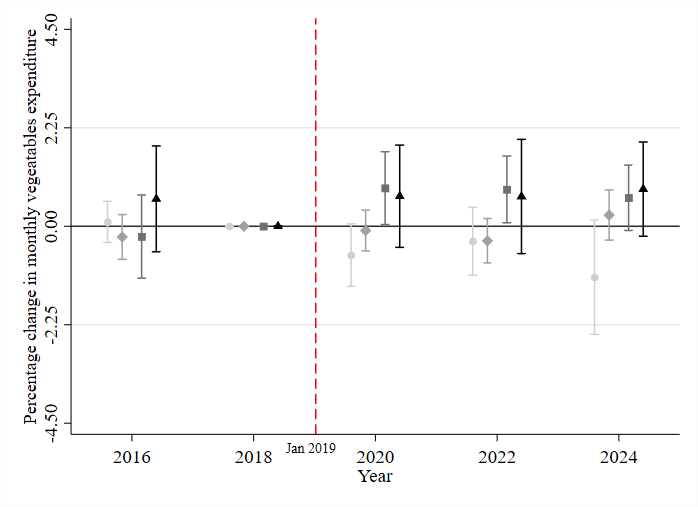}
  \caption{Vegetables}
\end{subfigure}
\hfill
\begin{subfigure}{.49\textwidth}
  \centering
  \includegraphics[width=1\linewidth]{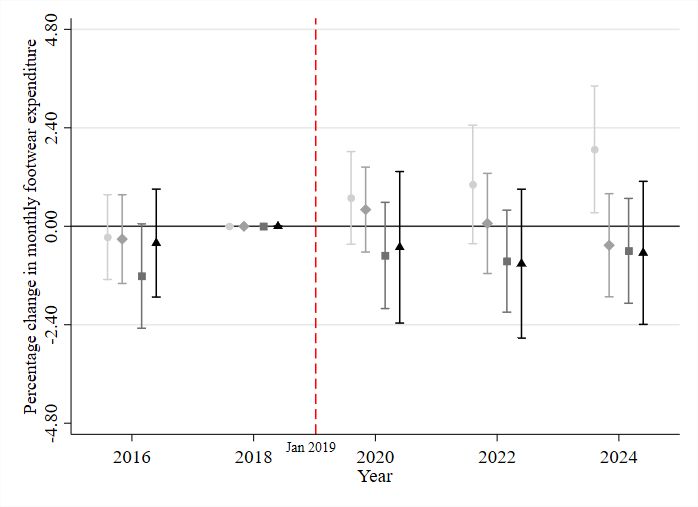}
  \caption{Footwear}
\end{subfigure}
\hfill
\fnote{\footnotesize \textit{Note:} Graph points show differences in outcomes between treatment and control groups relative to 2018 levels, between new eligible individuals with a contributory pension and the always eligible without a contributory pension. Estimates come from \autoref{event}. The vertical line depicts the date when the policy intervention was enacted (January 2019). This figure provides suggestive evidence on the parallel trend assumption underlying our DDD identification strategy. Data are from the 2016, 2018, 2020, 2022, and 2024 Household Income and Expenditure National Survey (ENIGH).}
\label{figConsumption}
\end{figure}

We begin by examining changes in household consumption patterns, which may have important implications for future health outcomes. Figure~\ref{figConsumption} reveals substantial heterogeneity across the income distribution. Among middle- and high-income households, the universalization of the pension program increased expenditures on cereals, meat, eggs, and vegetables. These changes are consistent with improvements in dietary quality and may contribute to better health outcomes in the medium and long run. In contrast, low-income households experienced reductions in the consumption of these goods, including root vegetables. Interestingly, these households increased expenditures on footwear, suggesting a reallocation of resources toward other forms of consumption. Whether these adjustments represent a temporary response to the reform or a more persistent change in household consumption behavior remains an open question for future research.

\begin{figure}[ht!]
\centering
\caption{Mechanism: Health, education and labor}
\begin{subfigure}{.49\textwidth}
  \centering
  \includegraphics[width=1\linewidth]{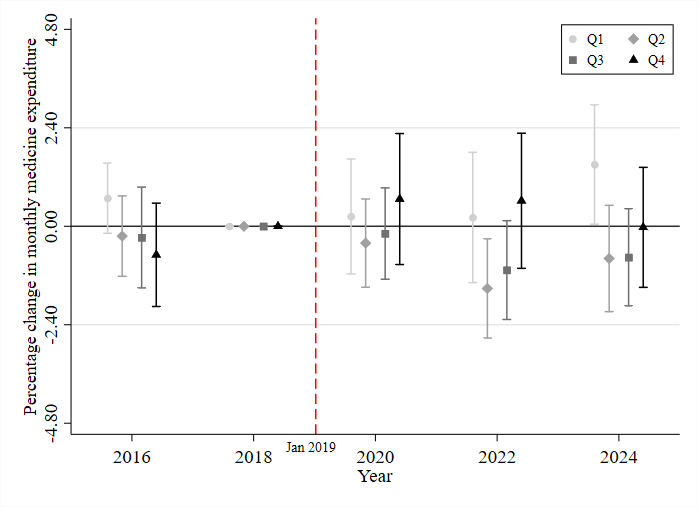}
  \caption{Medical expenses}
\end{subfigure}
\hfill
\begin{subfigure}{.49\textwidth}
  \centering
  \includegraphics[width=1\linewidth]{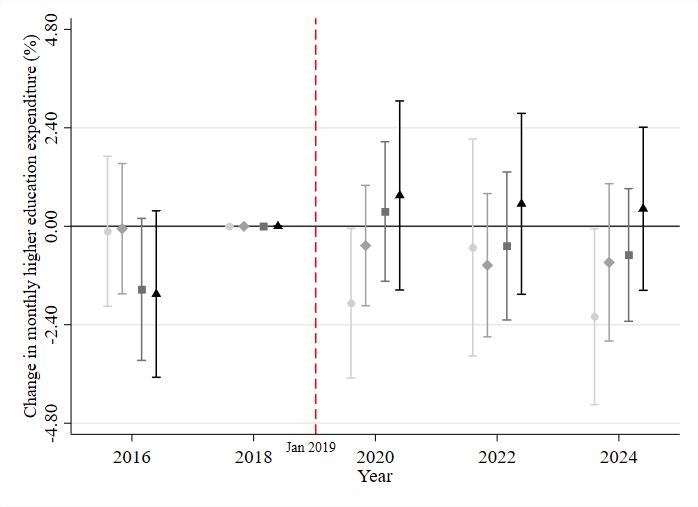}
  \caption{College expenses}
\end{subfigure}
\hfill
\begin{subfigure}{.49\textwidth}
  \centering
  \includegraphics[width=1\linewidth]{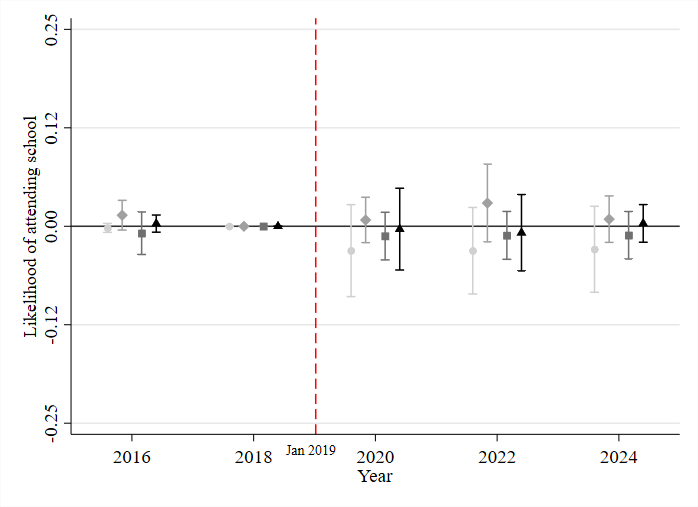}
  \caption{School attendance (K-12)}
\end{subfigure}
\hfill
\begin{subfigure}{.49\textwidth}
  \centering
  \includegraphics[width=1\linewidth]{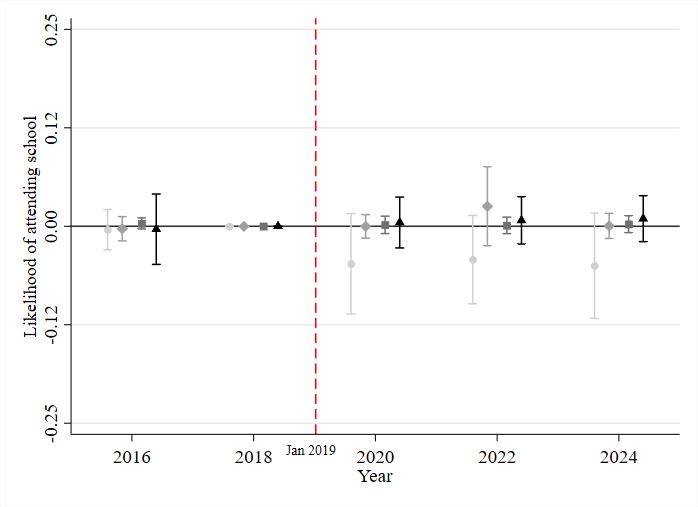}
  \caption{College attendance}
\end{subfigure}
\hfill
\begin{subfigure}{.49\textwidth}
  \centering
  \includegraphics[width=1\linewidth]{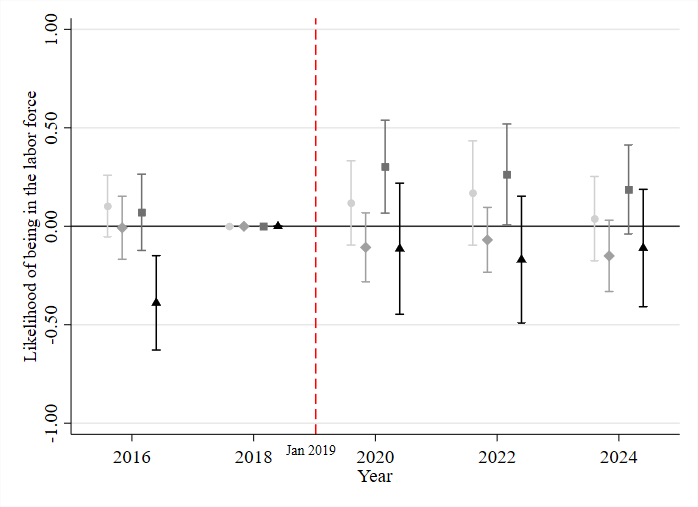}
  \caption{Labor force participation (6-18)}
\end{subfigure}
\hfill
\begin{subfigure}{.49\textwidth}
  \centering
  \includegraphics[width=1\linewidth]{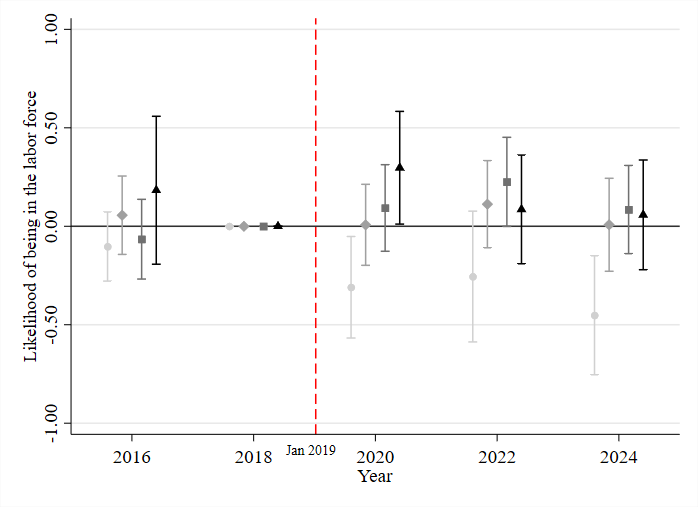}
  \caption{Labor force participation (19-30)}
\end{subfigure}
\hfill
\fnote{\footnotesize \textit{Note:} Graph points show differences in outcomes between treatment and control groups relative to 2018 levels, between new eligible individuals with a contributory pension and the always eligible without a contributory pension. Estimates come from \autoref{event}. The vertical line depicts the date when the policy intervention was enacted (January 2019). This figure provides suggestive evidence on the parallel trend assumption underlying our DDD identification strategy. Data are from the 2016, 2018, 2020, 2022, and 2024 Household Income and Expenditure National Survey (ENIGH).}
\label{figHealthEdu}
\end{figure}

We next examine two dimensions that are more directly related to human capital accumulation: health and education. Panel (a) of Figure~\ref{figHealthEdu} provides evidence of heterogeneous effects on health-related expenditures. Medical expenditures increased among households in the second and third quartiles of the income distribution, suggesting greater investment in health inputs. By contrast, the poorest households experienced declines in health expenditures, a pattern consistent with the broader evidence that the universalization generated adverse economic consequences among the most vulnerable elderly.

Educational outcomes display a similarly unequal pattern. Basic education expenditures remain largely unaffected across the income distribution. However, households in the lowest income quartile exhibit significant reductions in college-related expenditures and lower rates of school attendance among young household members. These effects are particularly concerning because they are not accompanied by greater labor force participation among the affected youth. On the contrary, the evidence suggests a reduction in labor market attachment, indicating that some young individuals may be withdrawing from both higher education and employment following the reform. This pattern is consistent with an increase in inactivity rather than a reallocation from schooling into work.

In contrast, young individuals living in middle- and high-income households appear to respond differently. For these groups, the universalization of the pension program is associated with stronger labor market participation and no evidence of reductions in educational investments. Taken together, these findings suggest that the effects of the reform extended beyond elderly beneficiaries and may have influenced the accumulation of human capital within the household. While middle- and high-income households appear to have used the additional resources to improve nutrition, health investments, and labor market opportunities, poorer households experienced outcomes that may have adverse implications for long-run human capital development.

\section{Robustness and heterogeneity}
\label{Heterogeneity}
\subsection{Robustness: Other social programs}
\label{Heterogeneity_Cohab}
A potential concern is that concurrent policy changes in Mexico may confound the estimated effects of the universalization of the Mexican Social Pension Program. The most prominent of these is the elimination of the conditional cash transfer program PROSPERA (formerly PROGRESA) in 2019 and its replacement with Becas Benito Juárez (BBJ). The same period also saw the introduction of Jóvenes Construyendo el Futuro (JCF), a large-scale training and employment program targeted at young adults. Because both programs primarily affected younger household members, they could potentially influence some of our outcomes through household-level spillovers. Previous research has documented that these policy changes affected the educational and labor market decisions of youth \citep{marquez2025}. Their direct relevance for the elderly, however, is limited. Prior to its elimination, PROSPERA only supported older adults who were not beneficiaries of the social pension program, and the amount transferred was substantially smaller than the pension benefit. Likewise, JCF targeted young individuals who were neither employed nor enrolled in school, making direct exposure among the elderly effectively nonexistent.

Nevertheless, indirect effects cannot be ruled out. If the replacement of PROSPERA with BBJ or the introduction of JCF altered household income or labor market opportunities for younger household members, elderly individuals living in cohabitation may have experienced economic spillovers. We define cohabitation as a living arrangement in which older adults reside with younger household members, such as children or grandchildren. Under this hypothesis, any confounding effects from these programs should be concentrated among cohabiting elderly rather than among those living alone.

\begin{figure}[ht!]
\centering
\caption{Robustness: Cohabitation}
\begin{subfigure}{.49\textwidth}
  \centering
  \includegraphics[width=1\linewidth]{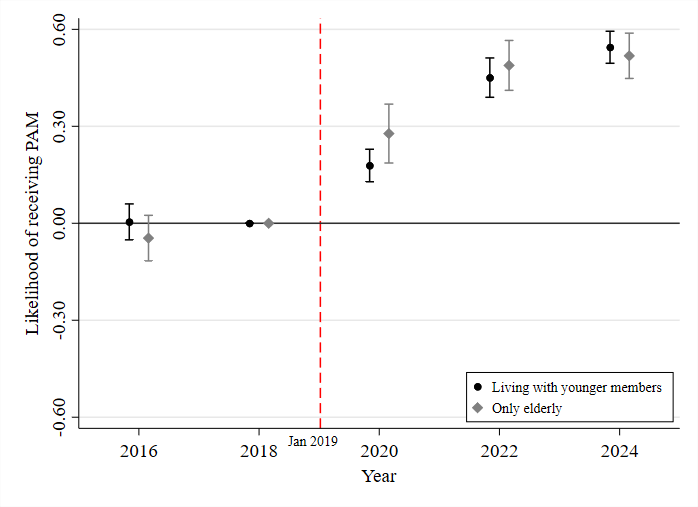}
  \caption{Program take-up}
\end{subfigure}
\hfill
\begin{subfigure}{.49\textwidth}
  \centering
  \includegraphics[width=1\linewidth]{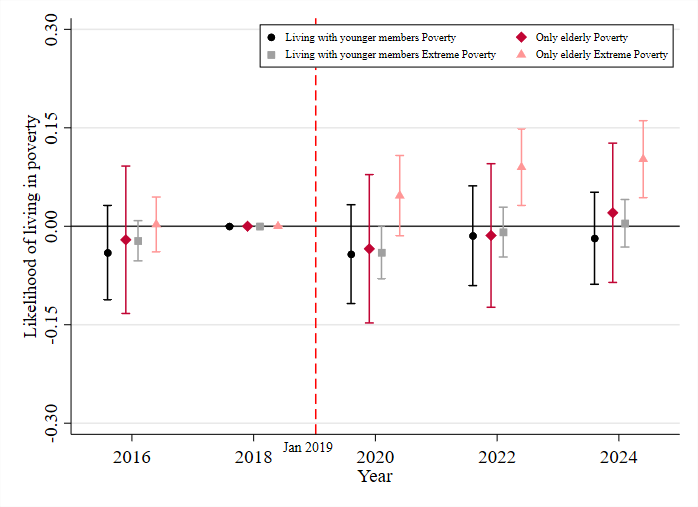}
  \caption{Poverty outcomes}
\end{subfigure}
\hfill
\begin{subfigure}{.49\textwidth}
  \centering
  \includegraphics[width=1\linewidth]{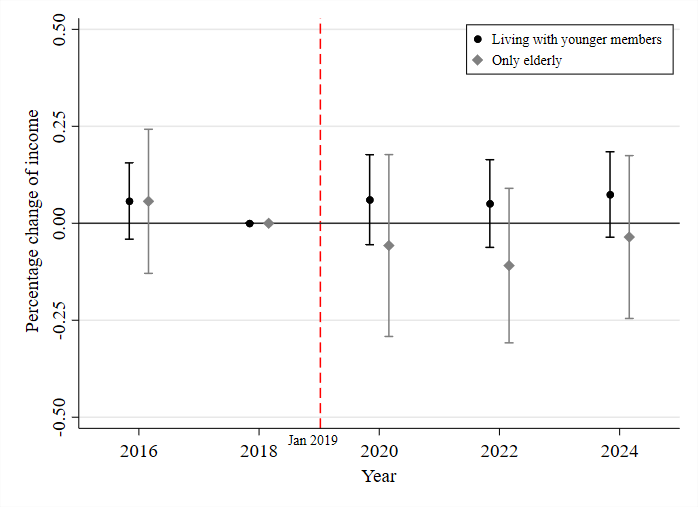}
  \caption{Ln(income)}
\end{subfigure}
\hfill
\begin{subfigure}{.49\textwidth}
  \centering
  \includegraphics[width=1\linewidth]{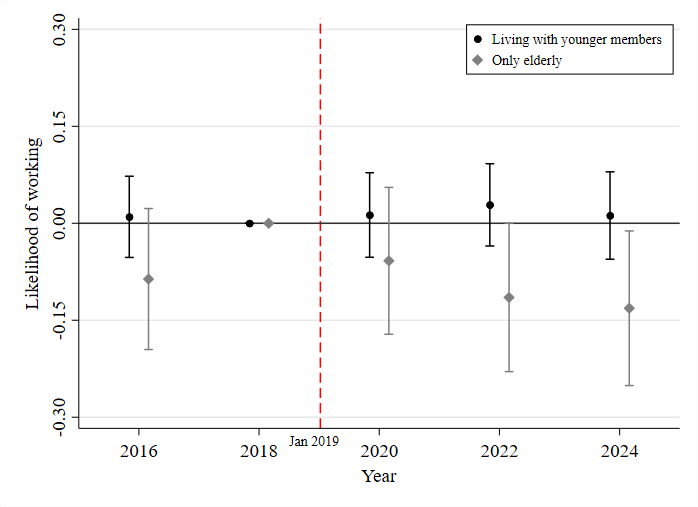}
  \caption{Work}
\end{subfigure}
\hfill
\fnote{\footnotesize \textit{Note:} Graph points show differences in outcomes between treatment and control groups relative to 2018 levels, between new eligible individuals with a contributory pension and the always eligible without a contributory pension. Estimates come from \autoref{event}. Each color corresponds to a separate regression conditioning on the cohabitation condition. Whiskers show 95\% confidence intervals. The vertical line depicts the date when the policy intervention was enacted (January 2019). This figure provides suggestive evidence on the parallel trend assumption underlying our DDD identification strategy. Data are from the 2016, 2018, 2020, 2022, and 2024 Household Income and Expenditure National Survey (ENIGH).}
\label{figHeterCohab}
\end{figure}

Panel (a) of \autoref{figHeterCohab} shows that take-up rates increased after the reform among both elderly individuals living alone and those living in cohabiting households, with no statistically significant differences between the two groups. This pattern provides little support for the hypothesis that changes in enrollment behavior were driven by within-household information spillovers or by efforts to compensate for the loss of PROSPERA transfers. More importantly, Panel (b) shows that the increase in extreme poverty is concentrated among elderly individuals living alone, who are the least likely to have been affected by changes to PROSPERA, BBJ, or JCF through household spillovers. The same group also exhibits the largest decline in labor force participation. Taken together, these findings suggest that the observed increase in extreme poverty is more plausibly explained by behavioral responses to the universalization of the social pension program rather than by the contemporaneous introduction or elimination of other social programs.

\begin{figure}[ht!]
\centering
\caption{Robustness: Disability condition}
\begin{subfigure}{.49\textwidth}
  \centering
  \includegraphics[width=1\linewidth]{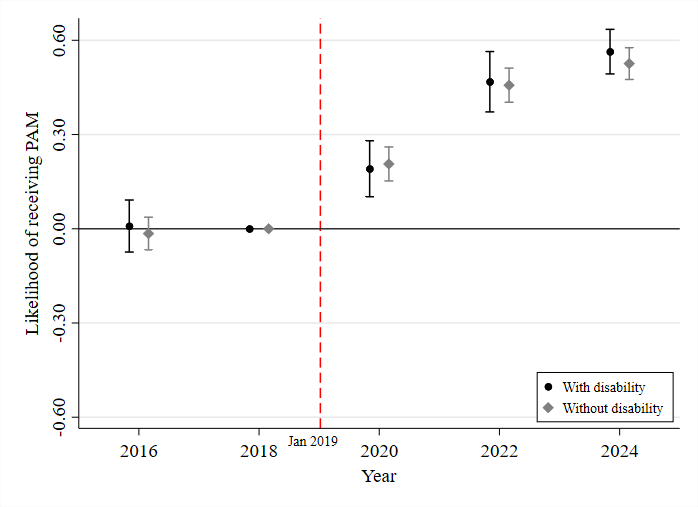}
  \caption{Program take-up}
\end{subfigure}
\hfill
\begin{subfigure}{.49\textwidth}
  \centering
  \includegraphics[width=1\linewidth]{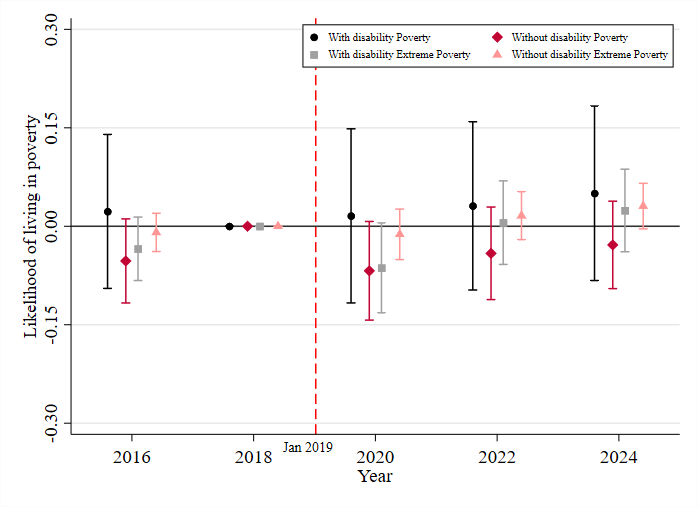}
  \caption{Poverty outcomes}
\end{subfigure}
\hfill
\begin{subfigure}{.49\textwidth}
  \centering
  \includegraphics[width=1\linewidth]{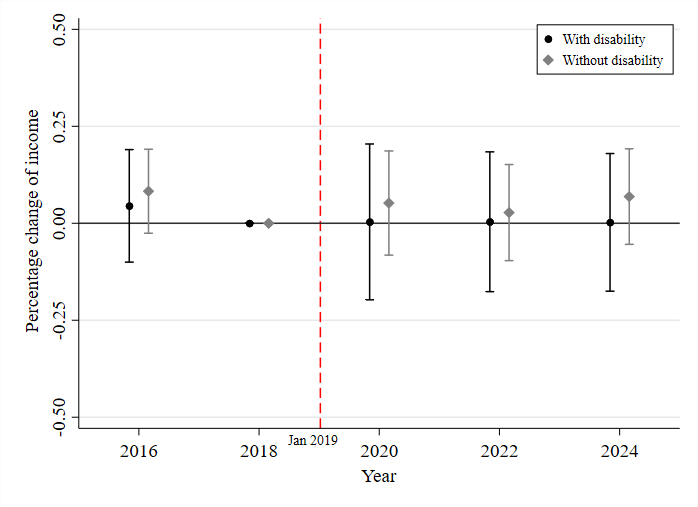}
  \caption{Ln(income)}
\end{subfigure}
\hfill
\begin{subfigure}{.49\textwidth}
  \centering
  \includegraphics[width=1\linewidth]{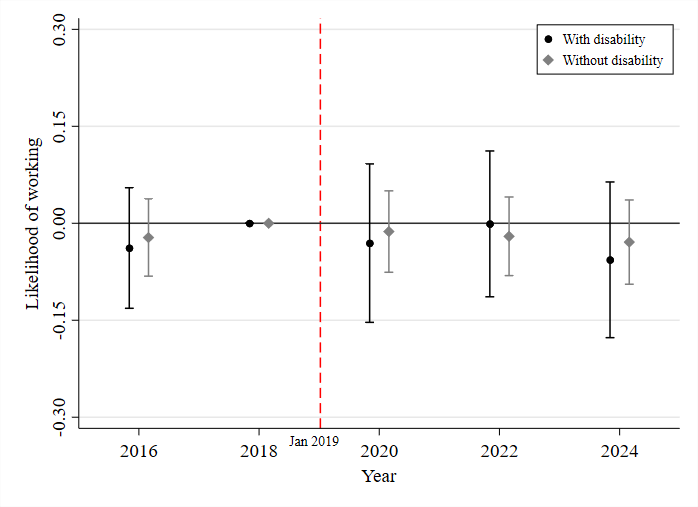}
  \caption{Work}
\end{subfigure}
\hfill
\fnote{\footnotesize \textit{Note:} Graph points show differences in outcomes between treatment and control groups relative to 2018 levels, between new eligible individuals with a contributory pension and the always eligible without a contributory pension. Estimates come from \autoref{event}. Each color corresponds to a separate regression conditioning on the disability condition. Whiskers show 95\% confidence intervals. The vertical line depicts the date when the policy intervention was enacted (January 2019). This figure provides suggestive evidence on the parallel trend assumption underlying our DDD identification strategy. Data are from the 2016, 2018, 2020, 2022, and 2024 Household Income and Expenditure National Survey (ENIGH).}
\label{figHeterDisability}
\end{figure}

Another concurrent policy was the creation in 2019 of the Pensión para el Bienestar de las Personas con Discapacidad Permanente (PBPDP), a transfer program for individuals with permanent disabilities. Although this program overlaps temporally with the universalization of the social pension, it is unlikely to bias our estimates. First, only 1.9\% of individuals in our sample belong to the relevant age and disability categories, according to the 2020 Mexican Population Census. Second, eligibility is determined by disability status and is orthogonal to our treatment definition. As a robustness check, we re-estimated the main specifications separately by disability status and found no statistically significant differences in treatment effects across groups (see \autoref{figHeterDisability}). These results further support the interpretation that our findings are driven by the universalization of the social pension program rather than by other contemporaneous policy interventions.

\subsection{Robustness: Minimum wage changes}
\label{Heterogeneity_MinWage}
Another possible confounder is the rapid increase in Mexico’s minimum wage starting in 2018. After years of stagnation, new legislation allowed for real increases: the minimum wage rose from 73.04 pesos/day in 2016 to 80.04 pesos/day in late 2017. In 2019, the reform introduced a geographically differentiated policy by creating the Zona Libre de la Frontera Norte (Northern Border Zone, NBZ), where the minimum wage jumped to 176.72 pesos/day, nearly twice the level in the rest of the country.

Between 2020 and 2024, nationwide minimum wages grew at annual rates between 15\% and 22\%. However, the NBZ experienced slightly slower growth, especially after 2023. These substantial hikes raise concerns that our estimated effects might reflect labor market responses to wage policy rather than to pension expansion.

We argue that such overlap is unlikely to bias our results for three main reasons. First, the labor supply effects we observe contradict what would be expected from rising minimum wages. If elderly individuals were responding to increased earnings potential, labor participation should rise or remain stable. Instead, we find consistent declines in employment, particularly among low-income seniors, suggesting the dominant force was increased access to non-labor income via PAM.

Second, Mexico’s contributory pensions are not indexed to the minimum wage. Rather, they are adjusted according to the Consumer Price Index (CPI), the Unidades de Medida y Actualización (UMA), or administrative formulas depending on the pension system (e.g., IMSS, ISSSTE, PEMEX, CFE, ISSFAM). Thus, increases in the minimum wage do not directly raise pension income, and any spillover effects would be minimal. Additionally, most contributory pensions are accessible well before age 65, so individuals in both treatment and control groups are exposed to these systems in comparable ways.

Third, our main results show no statistically significant effects of the reform on income, suggesting that wage-indexed spillovers (if any) did not materially alter measured earnings. The triple-differences design further controls for any common income shocks across groups.

To empirically verify these claims, we conduct a robustness check that exploits the regional wage policy. We compare treatment effects across two areas: the NBZ (with high minimum wages) and the rest of the country. If wage policy were driving our findings, we would expect significantly different outcomes in the NBZ, particularly in income and poverty.

\begin{figure}[ht!]
\centering
\caption{Robustness: Minimum wage}
\begin{subfigure}{.49\textwidth}
  \centering
  \includegraphics[width=1\linewidth]{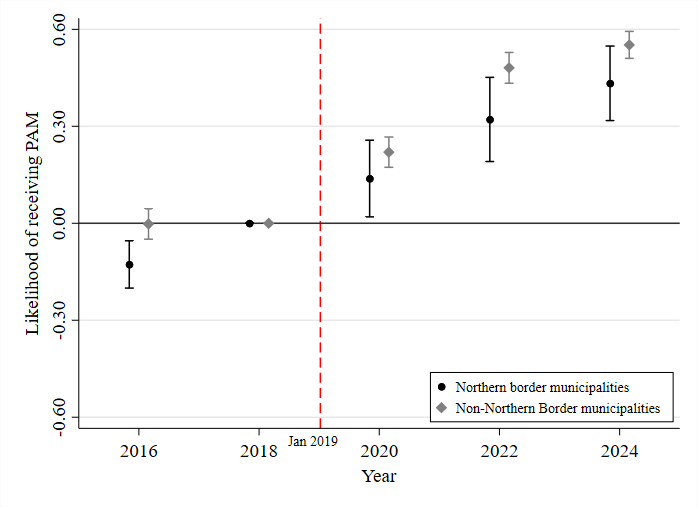}
  \caption{Program take-up}
\end{subfigure}
\hfill
\begin{subfigure}{.49\textwidth}
  \centering
  \includegraphics[width=1\linewidth]{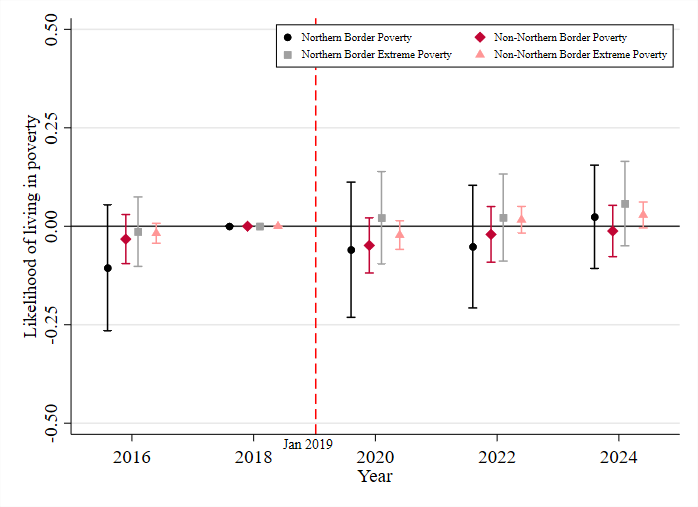}
  \caption{Poverty outcomes}
\end{subfigure}
\hfill
\begin{subfigure}{.49\textwidth}
  \centering
  \includegraphics[width=1\linewidth]{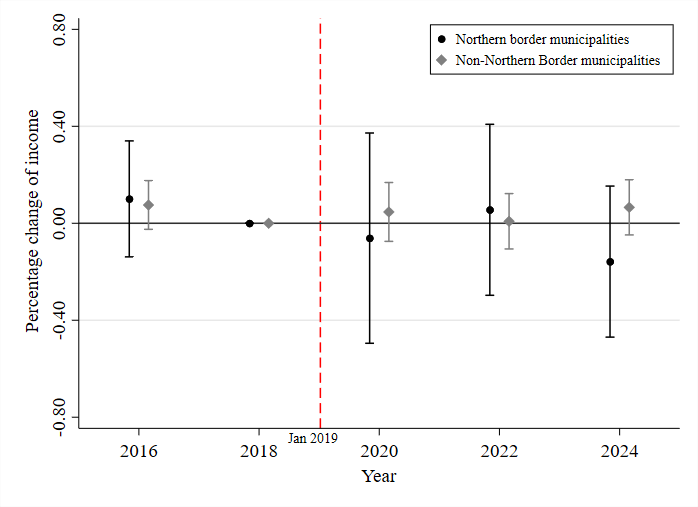}
  \caption{Ln(income)}
\end{subfigure}
\hfill
\begin{subfigure}{.49\textwidth}
  \centering
  \includegraphics[width=1\linewidth]{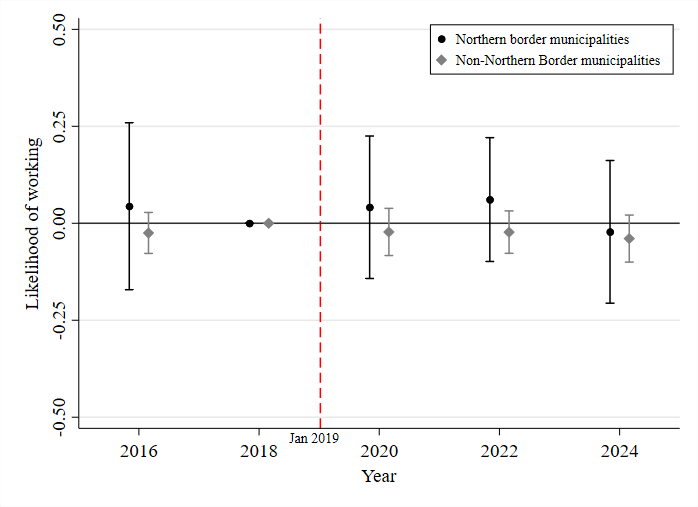}
  \caption{Work}
\end{subfigure}
\hfill
\fnote{\footnotesize \textit{Note:} Graph points show differences in outcomes between treatment and control groups relative to 2018 levels, between new eligible individuals with a contributory pension and the always eligible without a contributory pension. Estimates come from \autoref{event}. Each color corresponds to a separate regression conditioning on the geographical location: either the border zone or the non-border zone. Whiskers show 95\% confidence intervals. The vertical line depicts the date when the policy intervention was enacted (January 2019). This figure provides suggestive evidence on the parallel trend assumption underlying our DDD identification strategy. Data are from the 2016, 2018, 2020, 2022, and 2024 Household Income and Expenditure National Survey (ENIGH).}
\label{MinWage}
\end{figure}

\autoref{MinWage} shows no statistically significant differences in the treatment effects between zones across all outcomes. The only exception is a modestly lower take-up rate in the NBZ, likely driven by preexisting sociodemographic factors. Northern states historically have higher incomes and lower poverty rates, so fewer residents may have felt compelled to enroll. These results reinforce our conclusion that the observed labor and poverty dynamics are driven by PAM’s universalization, not by concurrent wage reforms.

\subsection{Robustness: Age cohort selection}
\label{Heterogeneity_Cohorts}
To address concerns that our estimates may reflect idiosyncratic features of the specific cohorts included in our main analysis, we conduct a robustness check using an alternative treatment group. Instead of focusing on individuals aged 68–70, we re-estimate our triple-differences specification using individuals aged 71–73 as the treatment group. The comparison group and other dimensions of variation (pension status and time) remain unchanged. This alternative cohort is also fully eligible for the PAM reform and shares similar characteristics with our baseline treatment group, but allows us to assess whether our findings generalize beyond the marginal age group closest to the cutoff.

\begin{figure}[ht!]
\centering
\caption{Robustness: Age cohort selection (external validity)}
\begin{subfigure}{.49\textwidth}
  \centering
  \includegraphics[width=1\linewidth]{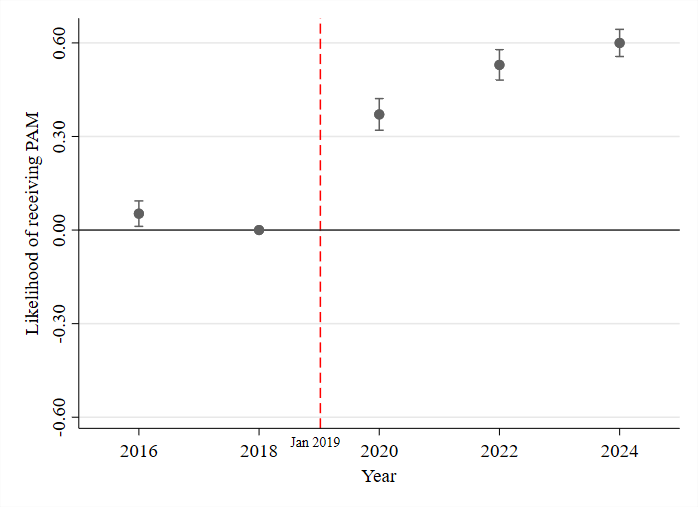}
  \caption{Program take-up}
\end{subfigure}
\hfill
\begin{subfigure}{.49\textwidth}
  \centering
  \includegraphics[width=1\linewidth]{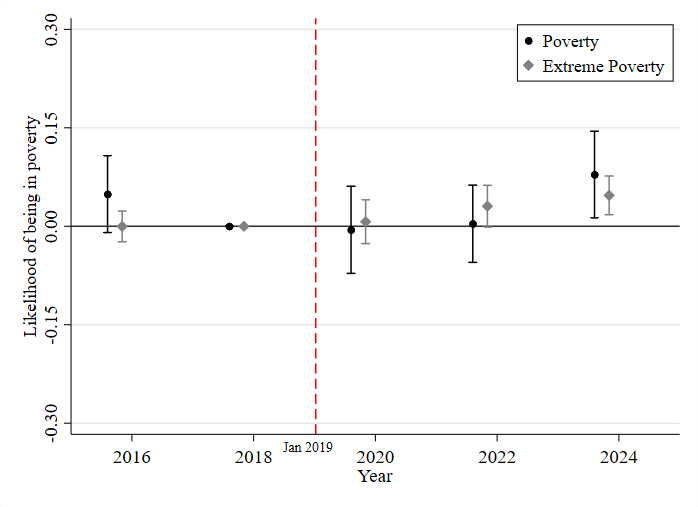}
  \caption{Poverty outcomes}
\end{subfigure}
\hfill
\begin{subfigure}{.49\textwidth}
  \centering
  \includegraphics[width=1\linewidth]{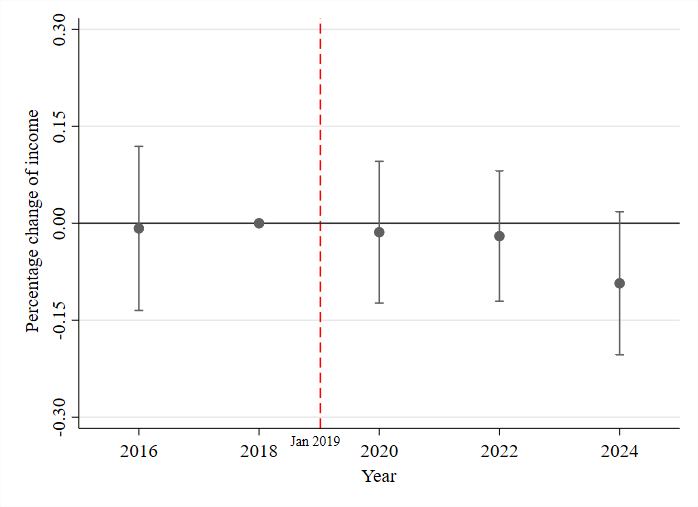}
  \caption{Ln(income)}
\end{subfigure}
\hfill
\begin{subfigure}{.49\textwidth}
  \centering
  \includegraphics[width=1\linewidth]{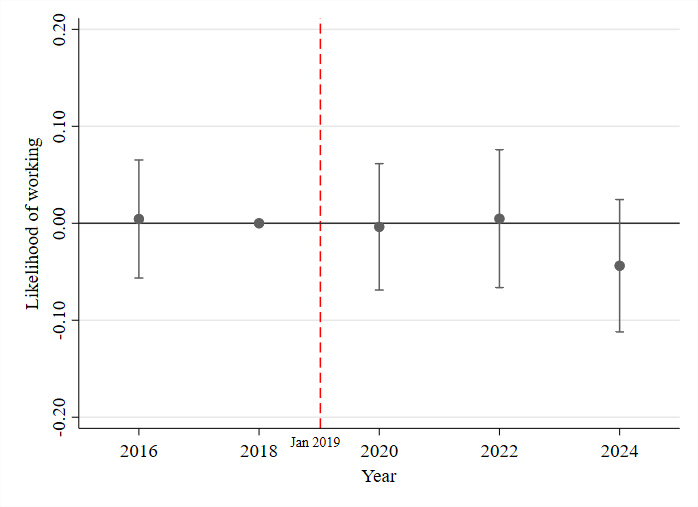}
  \caption{Work}
\end{subfigure}
\hfill
\fnote{\footnotesize \textit{Note:} Graph points show differences in outcomes between an alternative measure of treatment (71-73 year old elderly) and the control group (62-64  year old elderly) relative to 2018 levels, between new eligible individuals with a contributory pension and the always eligible without a contributory pension. Estimates come from \autoref{event}, with the adjustment on the definition of the treatment group. Whiskers show 95\% confidence intervals. The vertical line depicts the date when the policy intervention was enacted (January 2019). This figure provides suggestive evidence on the parallel trend assumption underlying our DDD identification strategy. Data are from the 2016, 2018, 2020, 2022, and 2024 Household Income and Expenditure National Survey (ENIGH).}
\label{Cohorts}
\end{figure}

The results support the robustness of our main findings. We continue to observe a sizable increase in program take-up and a rise in extreme poverty by 2024, consistent with our baseline results. In this specification, we also find a small but positive and statistically insignificant increase in overall poverty in 2024, aligning with the negative point estimate for income and a non-significant decline in employment (see \autoref{Cohorts}). These patterns reinforce our interpretation of adverse distributional consequences: the social pension may have prompted some elderly individuals to leave the labor market and rely solely on the transfer, despite its limited value relative to previous earnings. The consistency of results across age cohorts suggests that the effects of the universalization are not unique to the original treatment group and may extend to older eligible populations.

\subsection{Heterogeneity: Geographic heterogeneity}
\label{Heterogeneity_Geography}
\begin{figure}[ht!]
\centering
\caption{Heterogeneity: Geographical context}
\begin{subfigure}{.49\textwidth}
  \centering
  \includegraphics[width=1\linewidth]{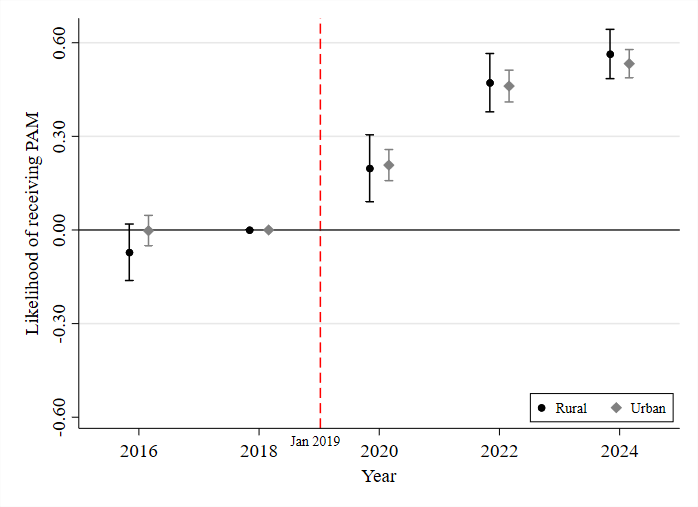}
  \caption{Program take-up}
\end{subfigure}
\hfill
\begin{subfigure}{.49\textwidth}
  \centering
  \includegraphics[width=1\linewidth]{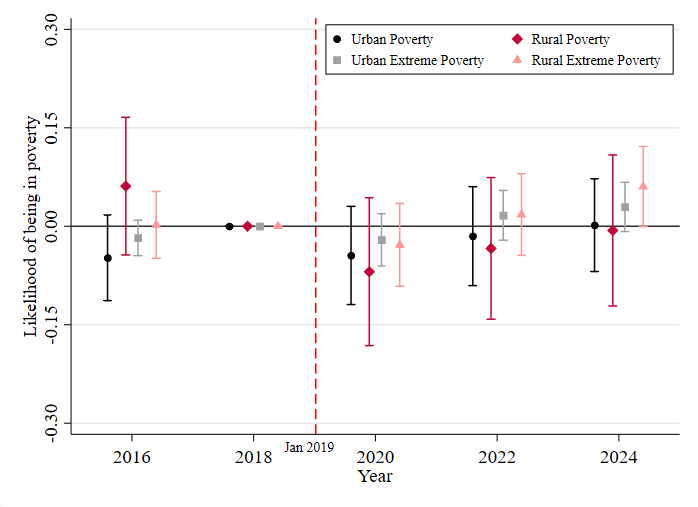}
  \caption{Poverty outcomes}
\end{subfigure}
\hfill
\begin{subfigure}{.49\textwidth}
  \centering
  \includegraphics[width=1\linewidth]{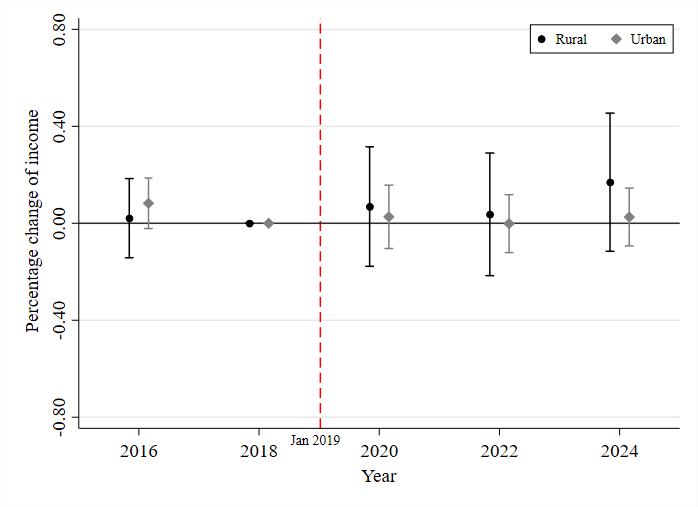}
  \caption{Ln(income)}
\end{subfigure}
\hfill
\begin{subfigure}{.49\textwidth}
  \centering
  \includegraphics[width=1\linewidth]{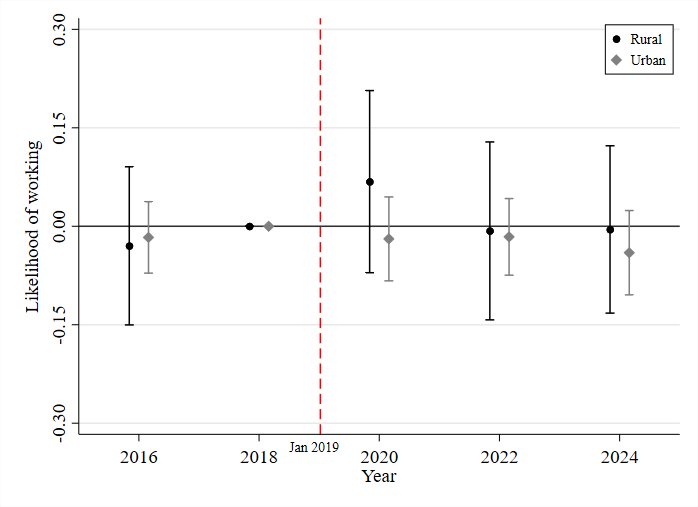}
  \caption{Work}
\end{subfigure}
\hfill
\fnote{\footnotesize \textit{Note:} Graph points show differences in outcomes between treatment and control groups relative to 2018 levels, between new eligible individuals with a contributory pension and the always eligible without a contributory pension. Estimates come from \autoref{event}. Each color corresponds to a separate regression conditioning on the geographical context: rural or urban. Whiskers show 95\% confidence intervals. The vertical line depicts the date when the policy intervention was enacted (January 2019). This figure provides suggestive evidence on the parallel trend assumption underlying our DDD identification strategy. Data are from the 2016, 2018, 2020, 2022, and 2024 Household Income and Expenditure National Survey (ENIGH).}
\label{figHeterRural}
\end{figure}

We also explore whether the effects of the reform varied by urban and rural status. \autoref{figHeterRural} shows no statistically significant differences in program take-up across rural and urban areas, suggesting that the expansion of PAM reached both contexts relatively evenly.

However, differences emerge in poverty outcomes. The increase in extreme poverty observed after 2022 is concentrated in rural areas. This may reflect the greater dependence of rural elderly on informal or subsistence labor, which may have been partially displaced by pension income. Additionally, rural households may face higher transaction costs in accessing complementary income sources, markets, or public services, making it harder to substitute lost labor income.

\subsection{Heterogeneity: Ethnicity}
\label{Heterogeneity_Ethnicity}
\begin{figure}[ht!]
\centering
\caption{Heterogeneity: Ethnicity}
\begin{subfigure}{.49\textwidth}
  \centering
  \includegraphics[width=1\linewidth]{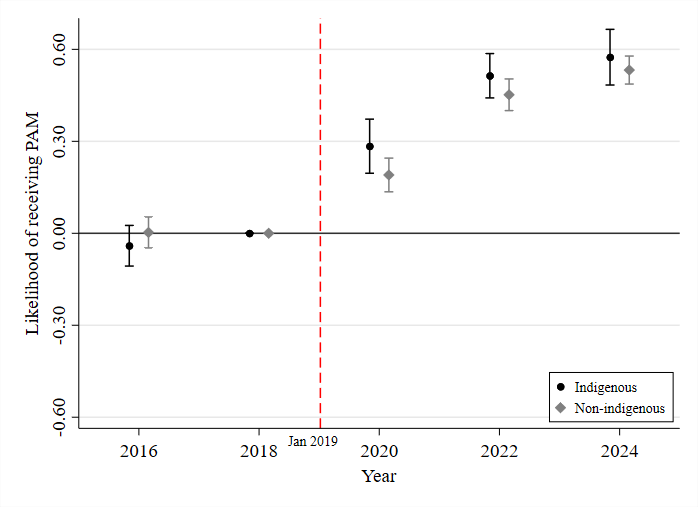}
  \caption{Program take-up}
\end{subfigure}
\hfill
\begin{subfigure}{.49\textwidth}
  \centering
  \includegraphics[width=1\linewidth]{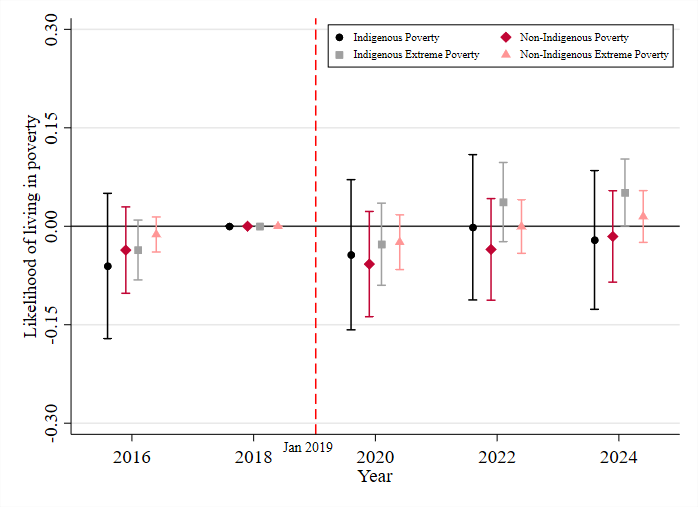}
  \caption{Poverty outcomes}
\end{subfigure}
\hfill
\begin{subfigure}{.49\textwidth}
  \centering
  \includegraphics[width=1\linewidth]{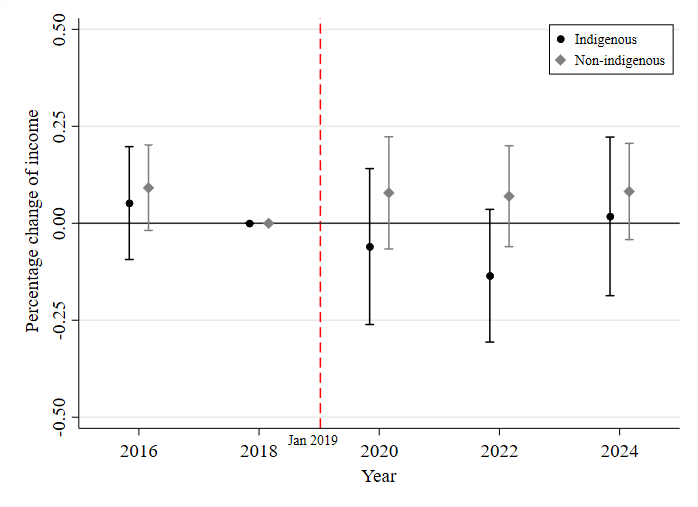}
  \caption{Ln(income)}
\end{subfigure}
\hfill
\begin{subfigure}{.49\textwidth}
  \centering
  \includegraphics[width=1\linewidth]{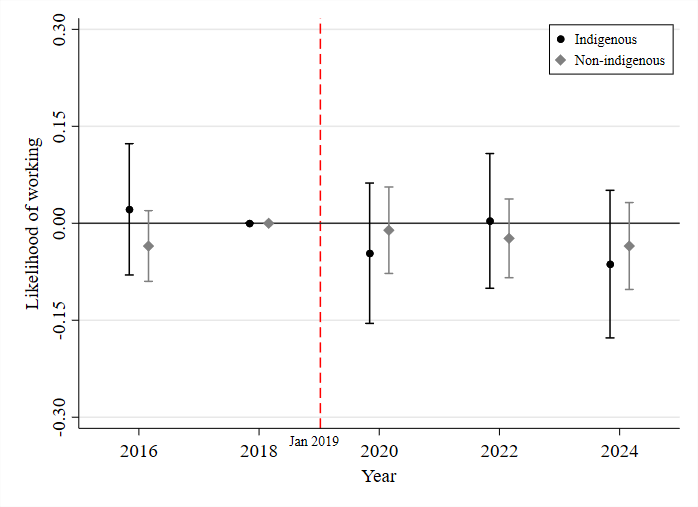}
  \caption{Work}
\end{subfigure}
\hfill
\fnote{\footnotesize \textit{Note:} Graph points show differences in outcomes between treatment and control groups relative to 2018 levels, between new eligible individuals with a contributory pension and the always eligible without a contributory pension. Estimates come from \autoref{event}. Each color corresponds to a separate regression conditioning on ethnicity. Whiskers show 95\% confidence intervals. The vertical line depicts the date when the policy intervention was enacted (January 2019). This figure provides suggestive evidence on the parallel trend assumption underlying our DDD identification strategy. Data are from the 2016, 2018, 2020, 2022, and 2024 Household Income and Expenditure National Survey (ENIGH).}
\label{figHeterIndig}
\end{figure}

Lastly, we examine heterogeneity by ethnicity. \autoref{figHeterIndig} shows that take-up rates among Indigenous individuals were consistently higher than those of non-Indigenous peers, especially in 2020. This is consistent with the implementation strategy of the 2019 expansion, which prioritized Indigenous communities during early rollout.

Panel (b) reveals that the post-reform increase in extreme poverty is disproportionately concentrated among Indigenous elderly. While the estimated effects on income are not statistically significant, the point estimates suggest a decline in labor activity. This is consistent with our broader interpretation that the pension induced partial retirement, particularly among economically vulnerable groups for whom the transfer was insufficient to compensate for lost labor income.

These findings underscore the importance of social and structural context (cohabitation, geographic remoteness, and Indigenous identity) in shaping the impacts of universal pensions. While the reform expanded access, it also introduced risks for subpopulations who may reduce work effort without adequate income replacement.

\section*{Conclusion and discussion}
\label{Conclusion}
This paper evaluates the effects of the 2019 expansion of Mexico’s Social Pension Program, one of the largest social protection reforms implemented in the country in recent decades. The reform simultaneously increased benefit generosity and expanded eligibility to individuals who had previously been excluded because they received contributory pensions, representing the first stage of a broader transition toward universal coverage. Using nationally representative household survey data and a combination of difference-in-differences and triple-differences strategies, we estimate the effects of these changes on poverty, labor supply, household behavior, and human capital investments.

Our findings reveal that the reform substantially increased program participation, largely through the expansion of eligibility. The combined effect of higher transfers and broader coverage generated meaningful reductions in both poverty and extreme poverty, primarily through increases in household income. However, once we isolate the universalization component of the reform, the results become considerably more nuanced. We find no evidence that universalization reduced poverty and, if anything, it contributed to a modest increase in extreme poverty by 2024.

The evidence is consistent with the labor-leisure mechanism developed in the theoretical framework. The pension increased non-labor income and appears to have induced some low-income elderly individuals to reduce labor supply. Because the transfer was insufficient to fully compensate for forgone labor earnings, the resulting decline in employment may have increased economic vulnerability among a subset of beneficiaries. These effects are particularly evident among elderly individuals living alone, in rural areas, and among Indigenous populations.

The analysis also reveals important household-level spillovers. Universalization affected consumption patterns, health expenditures, educational investments, and labor market outcomes among younger household members. While middle- and higher-income households generally used the additional resources to improve consumption and health-related expenditures, poorer households exhibited reductions in educational spending and school attendance at the college level. These findings suggest that the consequences of social pension reforms extend beyond direct beneficiaries and may influence the accumulation of human capital within multigenerational households.

At the same time, we find that a substantial share of higher-income elderly individuals enrolled in the program without experiencing significant changes in labor supply, income, or asset accumulation. This pattern is consistent with the principles of universal program design but highlights an important policy trade-off. Universalization can reduce exclusion and simplify access, particularly for disadvantaged populations, yet it also expands public expenditures toward groups whose welfare may be relatively unaffected by the transfer.

One possible policy response would be to preserve universal access while incorporating greater progressivity through the tax system. For example, making pension benefits taxable under the standard income tax schedule could allow governments to recover part of the transfer from higher-income beneficiaries without reintroducing administrative barriers or eligibility screens that may exclude vulnerable individuals.

More broadly, our findings suggest that evaluations of social pensions should look beyond immediate poverty outcomes and consider behavioral responses, household spillovers, and the distinction between expanding coverage and increasing benefit generosity. The Mexican experience illustrates that universalization can successfully broaden access to social protection, but its welfare consequences may differ substantially from those generated by larger transfers alone. Understanding these trade-offs is essential for designing equitable and fiscally sustainable pension systems in aging middle-income economies.

\section*{Acknowledgements}
We are grateful for the valuable comments of Edwin Van Gameren, Ornella Darova, Pablo Peña, Pablo Troncoso, Pia Orrenius, Raymundo Campos, Srinivasan Vasudevan, and the seminar participants at El Colegio de Mexico, the Southern Economic Association, and the Western Economic Association annual conferences.

\newpage{}

\singlespacing
\bibliography{ref_Univer}

\newpage{}
\section*{Figures and Tables}


\newpage

\begin{landscape}
\begin{center}
\input{tabSS.tex}
\end{center}
\end{landscape}
\begin{center}
\input{tab2.tex}
\end{center}

\end{document}

%% file: tabSS.tex
\begin{table}[ht!]\footnotesize
\def\sym#1{\ifmmode^{#1}\else\(^{#1}\)\fi}
\centering
\begin{threeparttable}
\caption{Summary statistics}
\label{ss}
\begin{tabular}{l*{12}{c}}
\toprule\midrule
& \multicolumn{5}{c}{Without a contributory pension} & &\multicolumn{5}{c}{With a contributory pension} &  \\\cmidrule{2-6}\cmidrule{8-12} 
& \multicolumn{2}{c}{Before} & &\multicolumn{2}{c}{After} & & \multicolumn{2}{c}{Before} & &\multicolumn{2}{c}{After} & \multicolumn{1}{c}{DDD} \\\cmidrule{2-3}\cmidrule{5-6} \cmidrule{8-9}\cmidrule{11-12}
& \multicolumn{1}{c}{Control} & \multicolumn{1}{c}{Treatment} & & \multicolumn{1}{c}{Control} & \multicolumn{1}{c}{Treatment} & & \multicolumn{1}{c}{Control} & \multicolumn{1}{c}{Treatment} & & \multicolumn{1}{c}{Control} & \multicolumn{1}{c}{Treatment} & \\ 
& \multicolumn{1}{c}{Group} & \multicolumn{1}{c}{Group} & & \multicolumn{1}{c}{Group} & \multicolumn{1}{c}{Group} & & \multicolumn{1}{c}{Group} & \multicolumn{1}{c}{Group} & & \multicolumn{1}{c}{Group} & \multicolumn{1}{c}{Group} & \\\cmidrule{1-13}
\multicolumn{13}{l}{\textit{Panel A: Outcome variables}} \\
[0.5em]
PAM take-up (\%) & 0.02 & 0.57 & &0.01 & 0.73 & & 0.00 & 0.08 & &0.02 & 0.68 & 0.42\sym{***}\\
 & (0.02) & (0.02) & &(0.01) & (0.01) & & (0.00) & (0.00) & &(0.02) & (0.02) & (0.02)\\
Ln(Income) & 7.96 & 7.86 & & 8.34 & 8.30 & & 8.59 & 8.54 & & 8.91 & 8.96 & 0.02\\
 & (0.03) & (0.03) & &(0.02) & (0.02) & & (0.03) & (0.03) & &(0.02) & (0.02) & (0.06)\\
Poverty (\%) & 0.51 & 0.56 & &0.44 & 0.45 & & 0.21 & 0.26 & &0.17 & 0.16 & -0.01\\
 & (0.01) & (0.01) & &(0.01) & (0.01) & & (0.01) & (0.01) & &(0.01) & (0.01) & (0.03)\\
Extreme Poverty & 0.16 & 0.19 & &0.13 & 0.13 & & 0.02 & 0.02 & &0.02 & 0.02 & 0.02\\
 & (0.01) & (0.01) & &(0.01) & (0.01) & & (0.00) & (0.00) & &(0.00) & (0.00) & (0.01)\\
Employment (\%) & 0.61 & 0.46 & &0.60 & 0.44 & & 0.28 & 0.23 & &0.23 & 0.18 & -0.00\\
 & (0.01) & (0.01) & &(0.01) & (0.01) & & (0.01) & (0.01) & &(0.01) & (0.01) & (0.02)\\
\cmidrule{2-13}
\multicolumn{13}{l}{\textit{Panel B: Control variables}}\\
[0.5em]
Female (\%) & 0.57 & 0.60 & &0.58 & 0.61 & & 0.41 & 0.40 & &0.44 & 0.41 & -0.02\\
 & (0.01) & (0.01) & &(0.00) & (0.00) & & (0.02) & (0.02) & &(0.01) & (0.01) & (0.03)\\
Education (years) & 6.11 & 4.92 & &7.17 & 5.68 & & 10.25 & 9.31 & &10.98 & 10.11 & 0.37\\
 & (0.15) & (0.15) & &(0.13) & (0.13) & & (0.20) & (0.20) & &(0.12) & (0.12) & (0.35)\\
Indigenous (\%) & 0.33 & 0.34 & &0.32 & 0.33 & & 0.22 & 0.24 & &0.21 & 0.21 & -0.03\\
 & (0.01) & (0.01) & &(0.01) & (0.01) & & (0.02) & (0.02) & &(0.01) & (0.01) & (0.02)\\
Cohabitation (\%) & 0.83 & 0.67& & 0.82& 0.66& & 0.77& 0.65& & 0.78& 0.65& 0.01\\
 & (0.00) & (0.00) & &(0.00) & (0.00) & & (0.00) & (0.00) & &(0.00) & (0.00) & (0.03)\\
Rural (\%) & 0.26 & 0.28 & &0.25 & 0.28 & & 0.07 & 0.08 & &0.08 & 0.08 & -0.02\\
 & (0.02) & (0.02) & &(0.01) & (0.01) & & (0.01) & (0.01) & &(0.01) & (0.01) & (0.01)\\
\cmidrule{1-13}
Observations & 8,163& 5,504& & 17,851& 11,574& & 2,434& 2,232& & 5,858& 5,326& 58,897\\
\midrule\toprule
\end{tabular}
\begin{tablenotes}[flushleft]
{\begin{spacing}{1.0}
\footnotesize{}      
\item \textit{Note:} In 2019, PAM was universalized. The table summarizes the characteristics of the treatment and control groups, observed before and after the reform, for individuals with access to a contributory pension and those without. The treatment group comprises individuals aged 68–70 who became eligible for PAM as a result of the universalization. The control group comprises ineligible individuals aged 62–64. The table reports weighted means and proportions; standard errors clustered at the municipality level are shown in parentheses. The last column presents estimates from our triple-differences (DDD) specification (\autoref{DDD}). Data come from the 2016–2024 waves of ENIGH survey. \sym{*} \(p<0.10\), \sym{**} \(p<0.05\), \sym{***} \(p<0.01\)
\end{spacing}} 
\end{tablenotes}
\end{threeparttable}
\end{table}

%% file: tab2.tex
\begin{table}[H]\centering
\def\sym#1{\ifmmode^{#1}\else\(^{#1}\)\fi}
\caption{The effect of PAM universalization on well-being outcomes \label{tab2}}
\begin{threeparttable}
\begin{tabular}{l*{5}{c}}
\hline\hline
&\multicolumn{1}{c}{(1)}&\multicolumn{1}{c}{(2)}&\multicolumn{1}{c}{(3)}&\multicolumn{1}{c}{(4)}&\multicolumn{1}{c}{(5)}\\
& PAM & Ln(Income) & Poverty & Extreme & Employed \\
&  &  &  & Poverty & \\
\hline
\multicolumn{6}{l}{\textit{Panel A. Combined effects: Full sample of Mexican elder}}\\
[0.5em]
Treatment effect (DiD) &       0.291\sym{***}&       0.092\sym{***}&      -0.058\sym{***}&      -0.026\sym{***}&      -0.013         \\
                    &     (0.011)         &     (0.020)         &     (0.012)         &     (0.009)         &     (0.011)         \\
\cmidrule{2-6}
Observations        & 58,897          &  58,897         &  58,897         &  58,897          &  58,897         \\
Adjusted R\textsuperscript{2} &       0.524         &       0.370         &       0.190         &       0.087         &       0.155         \\
\hline
\multicolumn{6}{l}{\textit{Panel B. Combined effects: Subsample of elder without a contributory pension}}\\
[0.5em]
Treatment effect (DiD) &       0.159\sym{***}&       0.099\sym{***}&      -0.059\sym{***}&      -0.038\sym{***}&      -0.007         \\
                    &     (0.011)         &     (0.022)         &     (0.014)         &     (0.012)         &     (0.012)         \\
\cmidrule{2-6}
Observations &      43,067         &      43,067         &      43,067         &      43,067         &      43,067         \\
Adjusted R\textsuperscript{2} &       0.561         &       0.346         &       0.158         &       0.081         &       0.225         \\
\hline
\multicolumn{6}{l}{\textit{Panel C. Combined effects: Subsample of elder with a contributory pension}}\\
[0.5em]
Treatment effect (DiD) &       0.583\sym{***}&       0.079\sym{*}  &      -0.061\sym{***}&      -0.011\sym{*}  &      -0.013         \\
                    &     (0.015)         &     (0.045)         &     (0.023)         &     (0.006)         &     (0.018)         \\
\cmidrule{2-6}
Observations &      15,830         &      15,830         &      15,830         &      15,830         &      15,830         \\
Adjusted R\textsuperscript{2} &       0.552         &       0.264         &       0.131         &       0.017         &       0.075         \\
\hline
\multicolumn{6}{l}{\textit{Panel D. Isolating the universalization effects}}\\
[0.5em]
Treatment effect (DDD) &       0.421\sym{***}&      -0.020         &      -0.001         &       0.026\sym{*}  &      -0.013         \\
                    &     (0.018)         &     (0.052)         &     (0.027)         &     (0.014)         &     (0.023)         \\
\cmidrule{2-6}
Observations        & 58,897          &  58,897         &  58,897         &  58,897          &  58,897         \\
Adjusted R\textsuperscript{2} &       0.558         &       0.385         &       0.207         &       0.097         &       0.234         \\
\hline\hline
\end{tabular}
\begin{tablenotes}[flushleft]
\footnotesize{}      
\item \textit{Note:} Panel A of this table shows the combined effects of PAM expansion in 2019, which includes the increment in the cash transfer and the universalization among all elder in Mexico. Panel B, shows the results only among elder without a contributory pension. These estimates are obtained with \autoref{did_eq}. Panel C, shows similar results as Panel B, but among elder with a contributory pension. Panel D shows the effect of universalizing PAM in Mexico (C-B). These estimates are obtained with \autoref{DDD}. The data source is the ENIGH survey for rounds 2016-2024.   
Standard errors clustered at the sample unit level in parentheses.
\sym{*} \(p<0.10\), \sym{**} \(p<0.05\), \sym{***} \(p<0.01\)
    \end{tablenotes}
\end{threeparttable}
\end{table}